\documentclass[12pt]{article}

\usepackage[usenames,dvipsnames]{xcolor}

\usepackage{tikz}
\usetikzlibrary{matrix,arrows,positioning,shapes,snakes,fadings,decorations.pathmorphing,
decorations.pathreplacing,automata,shadows,decorations.markings,patterns,decorations.text}
\usepackage{jheppub}
\usepackage{amsmath,amssymb,euscript,array,mathrsfs,appendix,ctable,mathabx}
\usepackage{mathtools,float}
\usepackage{arydshln}
\usepackage{todonotes}
\usepackage{graphicx}
\usepackage{wrapfig}
\usepackage{enumitem}
\usepackage{multirow}
\usepackage{bm,upgreek}
\usepackage{stmaryrd}

\usepackage{caption}
\usepackage{subcaption}

\usepackage{amsthm}

\usepackage{lineno}

\makeatletter
\newsavebox{\@brx}
\newcommand{\llangle}[1][]{\savebox{\@brx}{\(\m@th{#1\langle}\)}%
  \mathopen{\copy\@brx\kern-0.6\wd\@brx\usebox{\@brx}}}
\newcommand{\rrangle}[1][]{\savebox{\@brx}{\(\m@th{#1\rangle}\)}%
  \mathclose{\copy\@brx\kern-0.6\wd\@brx\usebox{\@brx}}}
\makeatother

\newcommand*\circled[1]{{\footnotesize\tikz[baseline=(char.base)]{%
            \node[shape=circle,fill=black!20,draw,inner sep=1pt] (char) {#1};}}}

\newcommand \arXiv [1]{\href{http://arxiv.org/abs/#1}{\tt arXiv:#1}}

\usepackage{titlesec}
\titleformat{\subsection}[display]{\it}{}{0.1cm}{\vspace{-1.5cm}\begin{center}\thesubsection\hspace{0.2cm}}[\end{center}\vspace{-0.5cm}]

\newcommand{\NORM}[1]{\big|\hspace{-0.5mm}\big| #1\big|\hspace{-0.5mm}\big| }
\newcommand{\ket}[1]{|#1\rangle}
\newcommand{\bra}[1]{\langle#1|}
\newcommand{\tr}{\text{Tr}}
\newcommand{\ext}{\text{ext}}

\newcommand\cc{c}

\newcommand{\EQ}[1]{\begin{equation}\begin{split} #1
\end{split}\end{equation}}

\title{The Holographic Map of an Evaporating Black Hole}
\author{Zsolt Gyongyosi$^a$, Timothy J. Hollowood$^a$, S.~Prem Kumar$^a$, Andrea Legramandi$^{a,b,c}$ and Neil Talwar$^a$}
\affiliation{$^a$Department of Physics, Swansea University, Swansea, SA2 8PP, U.K.\\
$^b$Pitaevskii BEC Center, CNR-INO and Dipartimento di Fisica, Università di Trento, I-38123 Trento, Italy\\
$^c$ INFN-TIFPA, Trento Institute for Fundamental Physics and Applications, Trento, Italy}
\emailAdd{z.gyongyosi.2133547@swansea.ac.uk,t.hollowood@swansea.ac.uk,\\ s.p.kumar@swansea.ac.uk, andrea.legramandi@unitn.it,\\ n.talwar.2017429@swansea.ac.uk}
\abstract{
We construct a holographic map that takes the semi-classical state of an evaporating black hole and its Hawking radiation to a microscopic model that reflects the scrambling dynamics of the black hole. The microscopic model is given by a nested sequence of random unitaries, each one implementing a scrambling time step of the black hole evolution. Differently from other models, energy conservation and the thermal nature of the Hawking radiation are taken into account. 
We show that the QES formula follows for the entropy of multiple subsets of the radiation and black hole. We further show that a version of entanglement wedge reconstruction can be proved by computing suitable trace norms and quantum fidelities involving the action of a unitary on a subset of Hawking partners. 
If the Hawking partner is in an island, its unitary can be reconstructed by a unitary on the radiation. We also adopt a similar setup and analyse reconstruction of unitaries acting on an infalling system.
}

\notoc
\begin{document}

\maketitle

\newpage 

\tableofcontents 

\section{Introduction}

Black holes lie at the front line of the struggle to unify quantum mechanics with gravity. Recent progress is focused on how this struggle plays out at the level of effective theory in a gravitating system like a black hole.  In particular, the effective description involves techniques that have evolved over many years involving quantum field theory over a fixed background spacetime using semi-classical techniques. In a black hole geometry this leads to the emission of Hawking radiation and the apparent loss of unitarity \cite{Hawking:1974sw,Hawking:1976ra}.  On the other hand, there is a microscopic level of description, for example provided by string theory, in which a black hole is described as a quantum system with a large density of states given by the Bekenstein-Hawking (BH) entropy (see \cite{Bena:2022rna} for a review). 

\subsection{The holographic map}

Recent progress has shed light on how these two levels of description are related and how the information-loss paradox is resolved and unitarity is restored \cite{Penington:2019npb,Almheiri:2019psf,Penington:2019kki,Almheiri:2019qdq} (also see the  reviews \cite{Almheiri:2020cfm,Bousso:2022ntt}). A key ingredient is a map, the `holographic map', between the effective semi-classical description and the microscopic description
\EQ{
V:\quad {\cal H}^\text{sc}\rightarrow{\cal H}^\text{micro}\ .
\label{nat}
}
The idea of such a map between the semi-classical and microscopic descriptions naturally arises in holography where the semi-classical state describes the state of bulk gravitational theory while the microscopic state describes the non-gravitational CFT dual \cite{Hamilton:2006az,Heemskerk:2012mn,Bousso:2012mh,Czech:2012bh,Wall:2012uf,Headrick:2014cta}. It is becoming clear that such a map should apply more generally and specifically in spacetimes which are not asymptotically AdS, such as an evaporating black hole, where the radiation can escape the AdS bulk. The holographic map has been interpreted as the encoding map of a quantum error code and this synergy between the two subjects has been very fruitful and has led to a better understanding of entanglement wedge reconstruction \cite{Almheiri:2014lwa,Pastawski:2015qua,Jafferis:2015del,Dong:2016eik,Harlow:2016vwg,Hayden:2016cfa,Cotler:2017erl,Faulkner:2017vdd,Hayden:2017xed,Akers:2019wxj,Akers:2020pmf}. However, recent work \cite{Akers:2021fut,Akers:2022qdl}\footnote{See also \cite{Maxfield:2022sio,Kim:2022pfp} for recent developments.} has clarified certain details and in particular argued that, in the context of a black hole, it is an important feature that the map is not isometric, $V^\dagger V\neq1$. This means that the relation with the standard theory of quantum error correcting codes is not so compelling. The non-isometric nature of the map is actually very natural because as the black hole ages its Hilbert space becomes too small to accommodate all the Hawking partners of the previously emitted radiation and so something has to give. Another key insight of \cite{Akers:2022qdl} is that the map does not act on the radiation once it has dispersed away from the black hole. This clarifies certain statements that have been made about the radiation, in particular it is not possible to change the microscopic state of the black hole by making operations on the radiation however complicated: there is no long-range non-locality of this kind.

The purpose of this work is to construct the holographic map $V$ in a very simple microscopic model of black hole evaporation defined e.g. in \cite{Hayden:2018khn,Hollowood:2021lsw} but refined to take account of energy conservation leading to thermal states. The basic version of the model is the block random unitary model (BRU) of \cite{Akers:2022qdl}. A number of key features follow also for this more refined model:

\begin{enumerate}[label=\protect\circled{\arabic*}]
\item The semi-classical state of the radiation $\rho_R^\text{sc}$ is precisely the average of the microscopic state of the radiation $\rho_R$ over the quasi-random microscopic scrambling dynamics of the black hole.
\item Past the Page time the quasi-random fluctuations of the microscopic state $\rho_R$ overwhelm the state and it becomes very different from the semi-classical (Hawking) state $\rho_R^\text{sc}$.
\item The Quantum Extremal Surface (QES) formula \cite{Faulkner:2013ana,Engelhardt:2014gca} for the entropy of a generic number of radiation and black hole subsets is derived in the regime where the black hole is evaporating slowly \cite{Hollowood:2021nlo,Hollowood:2021lsw}.
\item Unitary actions on an infalling system can be reconstructed on the radiation after the Page time showing that the information of the infalling system has been teleported out of the black hole realizing the Hayden-Preskill `black hole as a mirror' scenario \cite{Hayden:2007cs}. 
\item There is a version of state-specific entanglement wedge reconstruction (of the type discussed in \cite{Akers:2022qdl}): local unitaries acting on the Hawking partners can be reconstructed as a unitary acting on the black hole before the Page time and on the radiation after the Page time. The discussion is extended for generic subsets of the Hawking radiation.
\end{enumerate}

Let us now put some flesh on the bones. At the semi-classical level, the state of a QFT in the black hole background consists of an entangled state between the outgoing Hawking radiation $R$ and their partner modes behind the horizon $\overline R$. The overall state is pure
\EQ{
\ket{\psi}=\Big\{\sum_J\lambda_J\ket{J}_R\otimes\ket{J}_{\overline R}\Big\}\otimes\ket{S}_F\in{\cal H}^{\text{sc}}\ .
\label{sit}
}
We have also included the possibility for infalling modes in the state $\ket{S}_F$, including the matter that collapsed to form the black hole. We will develop two models: (i) a simple one in which the Hilbert space of the radiation is taken to be finite dimensional and \eqref{sit} is the maximally entangled state $\lambda_J=1/\sqrt{d_R}$ and (ii) a more refined one for which the radiation and partners are in a thermofield double with a slowly varying temperature.

At the microscopic level, the black hole is described by a finite dimensional Hilbert space ${\cal H}_B$ whose dimension is exponential in the BH entropy $d_B=e^{S_\text{BH}}$. The black hole emits Hawking radiation and at the microscopic level we can write the state of a partly evaporated black hole and radiation as
\EQ{
\ket{\Psi}=\sum_J\lambda_J\ket{J}_R\otimes\ket{\Psi_J}_B\in{\cal H}^\text{micro}\ .
\label{sat}
}

The two states, the semi-classical $\ket{\psi}$ and the microscopic $\ket{\Psi}$ are related by the holographic map \eqref{nat} 
\EQ{
V:\quad {\cal H}_R\otimes{\cal H}_{\overline R}\otimes{\cal H}_F\to{\cal H}_R\otimes{\cal H}_B\ .
} 
It was argued in \cite{Akers:2022qdl} that the map should act trivially on $R$ because the outgoing radiation system is identical in both the semi-classical and microscopic descriptions. So $V$ actually only acts non-trivially as  ${\cal H}_{\overline R}\otimes{\cal H}_F\to{\cal H}_B$. This is natural because the Hawking partner modes $\overline R$ and the infalling modes $F$ are behind the horizon and so part of the black hole whose semi-classical geometry should emerge from the microscopic description. By comparing \eqref{sit} with \eqref{sat}, we have
\EQ{
V\ket{J}_{\overline R}\otimes\ket{S}_F=\ket{\Psi_J}_B\ .
\label{hix}
}
We leave the dependence on the infalling state implicit.

The way that Hawking's information loss paradox can be resolved now reveals itself. In Hawking's analysis, the state of the radiation is the reduced state, the maximally-mixed state in the basic model and a quasi-thermal state in the refined model
\EQ{
\rho^\text{sc}_R=\sum_J|\lambda_J|^2\ket{J}_R\bra{J}\ ,
}
since the partner mode states are orthonormal, ${}_{\overline R}\bra{J}K\rangle_{\overline R}=\delta_{JK}$. On the other hand, at the microscopic level,
\EQ{
\rho_R=\sum_{JK}\lambda_K\bar\lambda_J\xi_{KJ}\ket{K}_R\bra{J}\ ,\qquad \xi_{KJ}=\bra{\Psi_J}\Psi_K\rangle\ .
\label{rug}
}
The semi-classical state is devoid of internal correlations, information is lost and unitarity is violated. The microscopic state, on the other hand, can carry the correlations and repair unitarity if the inner products $\xi_{JK}$ are non-trivial. The fact that
\EQ{
\bra{\Psi_J}\Psi_K\rangle\neq\delta_{JK}\ ,
\label{lol}
}
implies that the holographic map $V$ is non-isometric, a key insight in \cite{Akers:2022qdl}. It is the non-isometric nature of $V$ that allows information to escape out of the black hole in the correlations induced by the inner product \cite{Papadodimas:2013kwa}. Such a release of information would presumably be interpreted as being a non-local process to a semi-classical observer. This is a major insight but perhaps to be expected when spacetime geometry is an emergent concept.

For a black hole past its Page time, when $S_\text{rad}\gg S_\text{BH}$, one would expect the states $\ket{\Psi_J}$ to be far from orthogonal because there are order $e^{S_\text{rad}(R)}$ states in a much smaller $e^{S_\text{BH}}$ dimensional Hilbert space. Roughly speaking, as previously argued e.g. in \cite{Penington:2019kki,Stanford:2020wkf}, we find
\EQ{
\bra{\Psi_J}\Psi_K\rangle=\begin{cases}1+{\mathscr O}(e^{-S_\text{BH}}) & J=K\ ,\\ {\mathscr O}(e^{-S_\text{BH}/2}) & J\neq K\ ,\end{cases}
\label{zoo}
}
so the violation appears to be exponentially small $\sim e^{-1/G}$ in the semi-classical limit. This seems to suggest that the corrections coming from the microscopic theory will be small. However, if we write $\xi=I+Z$, then $Z$ is roughly-speaking a quasi-random Hermitian matrix whose elements are order $e^{-S_\text{BH}/2}$. It seems, therefore, that the effect of $Z$ would be very suppressed. However, if the dimension of the matrix $\sim e^{S_\text{rad}}$ is large then its eigenvalues can be expected to lie in a distribution between $\pm e^{(S_\text{rad}-S_\text{BH})/2}$. What this indicates is that the fluctuations in $Z$ could be expected to give rise to a radical change in the state of the radiation beyond the Page time when $S_\text{rad}\gg S_\text{BH}$ and a mechanism to ensure the unitarity of the evaporation. On the other hand, if we average the microscopic state over the quasi-random fluctuations $Z$ we recover the semi-classical state
\EQ{
\overline{\rho_R}=\rho_R^\text{sc}\ .
\label{tot2}
}
The fact that $\rho_R\neq\rho_R^\text{sc}$ means that if we were to attempt to interpret the microscopic state as a state on the semi-classical geometry, then in the near-horizon region it would not be the inertial vacuum and so we could expect there will be non-trivial energy and momentum as the horizon is approached \cite{Almheiri:2012rt}.

Another issue that is clarified by the fact that $V$ acts trivially on the radiation $R$ is, as already mentioned, that the state of black hole is completely invariant under any local action on the radiation. In more detail, the most general local action is obtained by coupling $R$ to an auxiliary system $M$ and having them interact. On the semi-classical state
\EQ{
\ket{\psi}\otimes\ket{\varnothing}_M\longrightarrow \sum_\alpha K_\alpha\ket{\psi}\otimes\ket{\alpha}_M\ ,
}
for some orthonormal states $\ket{\alpha}$ of $M$ and where the operators $K_\alpha$ act on $R$.
This defines a quantum channel acting on $R$ and unitarity implies that $K_\alpha$ are Krauss operators $\sum_\alpha K_\alpha^\dagger K_\alpha=1$. Mapping this to the microscopic state, and using the fact that $[V,K_\alpha]=0$, the reduced state on $B$, after $R$ and $M$ have interacted, is
\EQ{
\rho'_B=\sum_{\alpha}\tr_R\Big\{K_\alpha\ket{\Psi}\bra{\Psi}K_\alpha^\dagger\Big\}=
\tr_R\Big\{\ket{\Psi}\bra{\Psi}\sum_\alpha K_\alpha^\dagger K_\alpha\Big\}=\rho_B\ ,
}
so the state of the black hole is invariant.

\subsection{The QES formula}

One can quantitatively appreciate how $\rho_R$ differs from $\rho_R^\text{sc}$ by calculating their von Neumann entropies. The entropy of the semi-classical state $\rho_R^\text{sc}$, suitably regularized, is just the thermal entropy of Hawking radiation familiar from Hawking's calculation. The question is, how to calculate the entropy of the microscopic state $\rho_R$? This is where the QES, or generalized entropy, formula comes in \cite{Ryu:2006bv,Hubeny:2007xt,Hubeny:2013gba,Wall:2012uf,Faulkner:2013ana}. It relates the von Neumann entropy of the microscopic state $\rho_A$ reduced on some subsystem factor e.g.~$A=R$ or $B$, or some more specific subset of $R$ to the generalised entropy:
\EQ{
S(\rho_A)=\min S_\text{gen}(X_A),\qquad S_\text{gen}(X_A) = \frac{{\mathscr A}(X_A)}{4G}+S(\rho^\text{sc}_{{\cal W}(A)}).
\label{cut}
}
Here $\mathscr{A}(X_A)$ is the area of a codimension two surface $X_A$ in the gravitating region, called the Quantum Extremal Surface (QES), that bounds a region known as the entanglement wedge ${\cal W}(A)$ of $A$ in the gravitating region. The full entanglement wedge ${\cal W}(A)$ is given by appending $A \cap R$ to this region\footnote{More precisely, ${\cal W}(A)$ is the domain of dependence of this region.} and is determined by extremising the generalised entropy. Note that if $A$ is the radiation $R$, or some subset thereof, the entanglement wedge ${\cal W}(A)$ consists of $A$ and potentially also a region disconnected from $A$ i.e. ${\cal W}(A) = A \cup I$. The region $I$ is known as the `entanglement island', or `island' for short.

The formula \eqref{cut} is remarkable in several ways but principally because it allows one to calculate the entropy of the microscopic state $\rho_A$ using only semi-classical techniques even when the details of the microscopic theory are not known. It does this by implicitly averaging over the complex chaotic microscopic dynamics of the black hole in the way familiar from statistical mechanics. More precisely, when computed in the semi-classical theory, we can think of the left hand side as being equal to the usual $n\to1$ limit of the R\'enyi entropies but averaged in the following way
\EQ{
S(\rho_A)=\lim_{n\to 1}\ \frac1{1-n}\log\overline{e^{(1-n)S^{(n)}(\rho_A)}}\ ,
}
with the average over a suitable ensemble that is a proxy for the underlying complex, chaotic microscopic dynamics. Just as in statistical mechanics, the conceptual idea is that the average captures the behaviour of a single typical microscopic state because, unlike the state itself, the R\'enyi entropies are self-averaging quantities.

For an evaporating black hole, the QES are behind the horizon and when the evaporation is slow, which it is for most of the evaporation time apart from the final stage, the QES are very close behind the horizon. In fact the QES are completely determined within the scope of the slow evaporation approximation \cite{Hollowood:2020kvk,Hollowood:2021nlo,Hollowood:2021lsw}. Firstly, they have Kruskal-Szekeres (KS) coordinates related via
\EQ{
UV\thicksim\frac{\cc}{S_\text{BH}}\ll1\ ,
\label{yet}
}
where $\cc$ are the number of (massless) fields. In terms of Eddington-Finkelstein (EF) coordinates $(u,v)$,\footnote{The KS and EF coordinates are related by an approximately exponential map, $U = - \exp\left(-2\pi \int^u T(t) dt\right)$ and $V = \exp\left(2\pi \int^v T(t) dt\right)$, where $T(t)$ is the instantaneous temperature of the black hole.} this means
\EQ{
v=u-\Delta t_\text{scr}\ ,\qquad \Delta t_\text{scr}=\frac1{2\pi T}\log\frac{S_\text{BH}}\cc\ .
\label{vic}
}
The slow evaporation regime applies precisely when $S_\text{BH}\gg c$ so that the QES are pressed up against the horizon from within. The time shift above between the infalling and outgoing coordinates $\Delta t_\text{scr}$ is identified with the scrambling time of the black hole. This is time dependent but only changes  slowly as the black hole evaporates.

The second condition on the QES is that the outgoing EF coordinate of a QES $u_\text{QES}$ (inside the horizon) must be 
equal to the outgoing EF coordinate of one of the endpoints of the radiation $u_{\partial A}$ (outside the horizon)
\EQ{
\{u_\text{QES}\}\subset \{u_{\partial A}\}\ .
\label{eq:extr}
} 
This reduces the variation problem to a discrete minimization problem. 

When $A$ is a subset of the radiation and the entanglement wedge ${\cal W}(A)=A\cup I$, the second term in \eqref{cut} is just the thermal entropy\footnote{There is a common divergence associated with the end-points of $A$ at $\mathscr I^+$ which can be regularized. The divergences associated to end-points of $I$, on the other hand, are precisely cancelled by the divergences in the area term in \eqref{cut}.}
\EQ{
S(\rho^\text{sc}_{{\cal W}(A)})\approx S_\text{rad}(A\ominus\tilde I)=\frac{\pi c}6 \int_{A\ominus\tilde I}T(u)\,du\ ,
\label{pil}
}
where $T(u)$ is the instantaneous temperature of the black hole as a function of the outgoing EF coordinate $u$ on $\mathscr I^+$. Here, $\tilde I$, the `island-in-the stream', is just the reflection of the island in the horizon and projected onto $\mathscr I^+$ \cite{Hollowood:2021nlo,Hollowood:2021lsw,Gyongyosi:2022vaf}
. So in terms of the outgoing EF coordinate $u$, $I$ and $\tilde I$ are equal, with the former outside the horizon and the latter inside. The symmetric difference in \eqref{pil} accounts for the fact that $I$ contains purifiers of the radiation. The first term in \eqref{cut} is then approximately equal to the Bekenstein-Hawking entropy $S_\text{BH}$ evaluated at EF outgoing coordinates of the QES $u_{\partial I}$. Hence, within the slow evaporation approximation, we can write the entropy as a discrete minimization problem
\EQ{
S(A)\approx\min_{I}\Big\{\sum_{u_{\partial I}}S_\text{BH}(u_{\partial I})+S_\text{rad}(A\ominus\tilde I)\Big\}\ .
\label{twm}
}
This formula can easily be adapted to the case when $A$ includes the black hole itself, $B\subset A$. One simply replaces $A$ by $A\cap R$ in the second term.\footnote{Then, if $R_N\not\in A$, the most recent emitted  interval of radiation, there must be a QES with a $u$ coordinate equal to the $u$ coordinate of the upper end-point of $R_N$, giving a contribution $S_\text{BH}(M_N)\equiv S_\text{BH}(M)$ to \eqref{twm}. On the other hand, if $R_N\in A$ then it must be that $R_N\subset I$. In the latter case, the connected subset of $I$ that includes $R_N$ is not strictly-speaking part of the island although it is in the entanglement wedge of $A$.} In section \ref{s4} we verify this formula in both the basic and refined models using the replica trick. We also find a simple formula \eqref{bom} for the island which relates the replica trick and the entanglement wedge in quite a direct way.

The paper is organized as follows. In section \ref{s2} we define two simple discrete models of a holographic map for an evaporating black hole. There is a basic and refined model. Compared with the basic model, the refined model has the nice features that the state of a small subsystem is thermal instead of maximally mixed and that the irreversiblity of evaporation is naturally incorporated (there is no need to add in ancilla qubits to mimic this effect). We also show that the average over the quasi-random unitary time evolution of the microscopic state is just the semi-classical state \eqref{tot2}. In section \ref{s4}, we compute the R\'enyi and von Neumann entropies of subsets of the radiation and black hole and derive the minimization problem  for the generalized entropy of a slowly evaporating black hole \eqref{twm}. In section \ref{s5}, we turn to the Hayden-Preskill scenario \cite{Hayden:2007cs} and consider when the action of a unitary on an infalling system be reconstructed (in a state-specific sense) on a subset of the radiation or black hole from a `decoupling argument'. We find the model reproduces the `black hole as a mirror' phenomenon and reconstruction is possible on the radiation when the black hole is past the Page time. This problem of reconstruction of operators acting on an infalling system was studied in the basic (or BRU) model and a random pairwise interaction model (which incorporates the fast scrambling nature of black holes) in \cite{Akers:2022qdl}. Our main contributions here are to study this problem in a model which generalises the basic model and also to consider when reconstruction is possible not just on the radiation or the black hole, but a subset thereof. In section \ref{s6} we consider when local operations, in the form of a quantum channel, acting on the Hawking partners can be reconstructed on a subset of the radiation or black hole. As expected, we find that reconstruction on the radiation is possible when the black hole is past the Page time. In section \ref{s7} we draw some conclusions. In appendix \ref{a2} we review the computation of certain thermodynamic quantities for free bosonic and fermionic fields, which is used in the refined model. In appendix \ref{a1} we provide a proof of the dominant saddles which contribute in the replica trick calculation in the basic model.

\section{The model}\label{s2}

\begin{figure}[h]
\begin{center}
\begin{tikzpicture} [scale=0.58]
\draw[fill = Plum!10!white,line width=0.6mm,rounded corners = 0.2cm] (-1,-1) rectangle (1,1);
\draw[-triangle 60,line width=0.4mm] (-3.8,0) -- (-1,0);
\node at (-2.5,-0.7) {$B_0=F_0$};
\draw[decorate,line width=0.4mm,decoration={snake,amplitude=0.03cm},-triangle 60] (1,0.5) -- (3.5,1.5) node[right,sloped] {$R_1$};
\draw[-triangle 60,line width=0.4mm] (1,-0.5)  -- (4,-0.5) node[sloped,midway,above] {$B_1$};
\node at (0,0) {$U_1$};
\begin{scope}[xshift=5cm,yshift=-1cm]
\draw[fill = Plum!10!white,line width=0.6mm,rounded corners = 0.2cm] (-1,-1) rectangle (1,1);
\draw[decorate,line width=0.4mm,decoration={snake,amplitude=0.03cm},-triangle 60] (1,0.5) -- (3.5,1.5) node[right,sloped] {$R_2$};
\draw[-triangle 60,line width=0.4mm] (1,-0.5)  -- (4,-0.5) node[sloped,midway,above] {$B_2$};
\node at (0,0) {$U_2$};
\draw[-triangle 60,line width=0.4mm] (-3.5,-1.5)  -- (-1,-0.5) node[sloped,midway,below] {$F_1$};
\end{scope}
\begin{scope}[xshift=16cm,yshift=-3cm]
\draw[fill = Plum!10!white,line width=0.6mm,rounded corners = 0.2cm] (-1,-1) rectangle (1,1);
\draw[-triangle 60,line width=0.4mm] (-3.5,0.5) -- (-1,0.5) node[sloped,midway,above] {$B_{N-1}$};
\draw[decorate,line width=0.4mm,decoration={snake,amplitude=0.03cm},-triangle 60] (1,0.5) -- (3.5,1.5) node[right,sloped] {$R_N$};
\draw[-triangle 60,line width=0.4mm] (1,-0.5)  -- (4,-0.5) node[below,midway] {$B\equiv B_N$};
\node at (0,0) {$U_N$};
\draw[-triangle 60,line width=0.4mm] (-3.5,-1.5)  -- (-1,-0.5) node[sloped,midway,below] {$F_{N-1}$};
\end{scope}
\draw[very thick,dotted] (9.8,-1.4) -- (11.8,-2.1);
\end{tikzpicture}
\caption{\footnotesize The model of black hole evaporation consisting of a sequence of random unitaries that mimic the scrambling microscopic dynamics. At each time step a small subsystem escapes as the Hawking radiation and there can be an infalling system. The time steps are of the order of the scrambling time of the black hole \eqref{vic} and so the model will appear to be continuous at time scales much larger than the scrambling time, including the Page time and the evaporation time.} 
\label{fig1}
\end{center}
\end{figure}
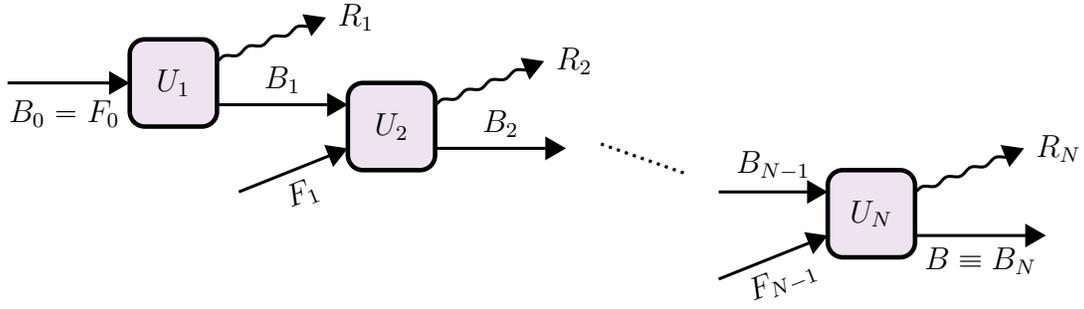

In the model, described in \cite{Hollowood:2021lsw}, the evaporation at the microscopic level is described by a series of discrete time steps identified with the scrambling time of the black hole \eqref{vic} shown in the figure \ref{fig1}. During the $p^\text{th}$ time step the state of the black hole evolves by a unitary $U_p$ which maps
\EQ{
U_p:\quad{\cal H}_{B_{p-1}}\otimes{\cal H}_{F_{p-1}}\longrightarrow {\cal H}_{B_p}\otimes{\cal H}_{R_p}\ .
} 
In the basic model, we have $d_{B_{p-1}}d_{F_{p-1}}=d_{B_p}d_{R_p}$, whereas in the refined model energy conservation is taken into account and $R_p$ is infinite dimensional. In this case, $U_p$ is an isometric embedding of a microcanonical energy window into ${\cal H}_{R_p}\otimes{\cal H}_{B_p}$, as we will describe later. After $N$ time steps, the state of the black hole and radiation is
\EQ{
\ket{\Psi(t_N)}=U_N\cdots U_2U_1\ket{S}_F \in {\cal H}_B\otimes  {\cal H}_R\ ,
\label{res}
}
where 
\EQ{
\ket{S}_F=\ket{s_0}_{F_0}\otimes\ket{s_1}_{F_1}\otimes\cdots\otimes\ket{s_{N-1}}_{F_{N-1}}\ ,
} 
describes the infalling matter that created the black hole $B_0\equiv F_0$ as well as matter that falls in during each time step $F_p$ as the black hole evaporates. The radiation is split into a temporal sequence of subsets $R=\bigcup_p R_p$. In the above, the remaining black hole is $B\equiv B_N$. A basis of states of the radiation consists of $\ket{J}_R$ where $J=\{j_1,\ldots j_N\}$ and each $j_p\in\{1,2,\ldots,d_{R_p}\}$ labels the states in the $p^\text{th}$ time step $R_p$. In particular, the microscopic states defined in \eqref{sat} are
\EQ{
\ket{\Psi_J}=\frac1{\lambda_J}\ {}_R\bra{J}U_N\cdots U_2U_1\ket{S}_F\in{\cal H}_B\ .
\label{pot}
}
Consequently the time evolution of the black hole in the model leads to a concrete expression for the holographic map,
\EQ{
V=\sum_J\frac1{\lambda_J}\ {}_{\overline R}\bra{J}\otimes{}_R\bra{J}U_N\cdots U_2U_1\ ,
\label{kil}
}
acting on ${\cal H}_{\overline R}\otimes{\cal H}_F$. In the basic model we take $\lambda_J = 1/\sqrt{d_R}$ and so the map has the form of a unitary followed by a post selection on the maximally-entangled state on ${\cal H}_{\overline R}\otimes{\cal H}_R$. In the refined model we take $\lambda_J$ to be given by \eqref{lambdaJ ref def}. In the refined model the map also has the form of a unitary followed by a post selection, however, we note that the post selection is not on the thermofield double state on ${\cal H}_{\overline R}\otimes{\cal H}_R$. In fact, in the refined model the sum in \eqref{kil} is not well defined and $V$ is only defined acting on suitable states such as $\ket{\psi}$.

The basic model \cite{Hollowood:2021lsw} is identified with the block random unitary model of \cite{Akers:2022qdl}. What is noteworthy is that, in the basic model, the post selection on the maximally entangled state, which manifests the non-isometric property of the map, is also the mechanism which allows information to be teleported out of the black hole.

\begin{figure}[ht]
\begin{center}
\begin{tikzpicture} [scale=1.4]
\draw[red!10,fill=red!10] (4.6,7.4) -- (1.2,4) -- (0.7,4) -- (4.35,7.65) -- cycle;
\draw[red!40] (6,6) -- (4,4);
\draw[red!40] (5.25,6.75)  -- (2.5,4);
\draw[red!40] (4.6,7.4) -- (1.2,4);
\draw[red!40] (4.35,7.65) -- (0.7,4);
\draw[red!40] (4.25,7.75) -- (0.5,4);
\draw[red!10,fill=red!10]  (1.8,7) -- (0,5.2) -- (0,4.7) -- (2.3,7);
\draw[red!40] (0.5,7)  -- (0,6.5);
\draw[red!40] (1.8,7) -- (0,5.2);
\draw[red!40] (2.3,7) -- (0,4.7);
\draw[red!40] (2.5,7) -- (0,4.5);
\draw[blue!10,fill=blue!10,opacity=0.5] (2,4) -- (0,6) -- (0,7) -- (3,4);
\draw[blue!40] (1,4) -- (0,5);
\draw[blue!40] (2,4) -- (0,6);
\draw[blue!40] (3,4) -- (0,7);
\draw[blue!40] (4,4) -- (1,7);
\draw[decorate,very thick,black!60,decoration={zigzag,segment length=1.5mm,amplitude=0.5mm}] (0,7) -- (3,7);
\draw[dash dot] (0,4) -- (0,7);
\draw[-] (4,4) -- (6,6) -- (3,9);
\draw[dash dot] (3,9) -- (3,7);
\draw[dashed] (3,7) -- (0,4);
\filldraw[black] (3,7) circle (1.5pt);
\node[rotate=-45] at (5,7.7) {$\mathscr I^+$};
%
\node[rotate=45] at (4.23,7.27) {\footnotesize $R_p$};
%
\node[rotate=-45] at (2.1,4.4) {\footnotesize $F_p$};
\node[rotate=45] at (1.75,6.7) {\footnotesize $\overline R_p$};
\node[rotate=45] at (0.6,4.5) {\footnotesize horizon};
\end{tikzpicture}
\caption{\footnotesize Subsets of Hawking modes $R_p$ are their entangled partners $\overline R_p$ behind the horizon and infalling modes $F_p$. The Hawking modes propagate out to null infinity $\mathscr I^+$. Each $R_p$ and $F_p$ lasts for a  scrambling time.}
\label{fig3} 
\end{center}
\end{figure}
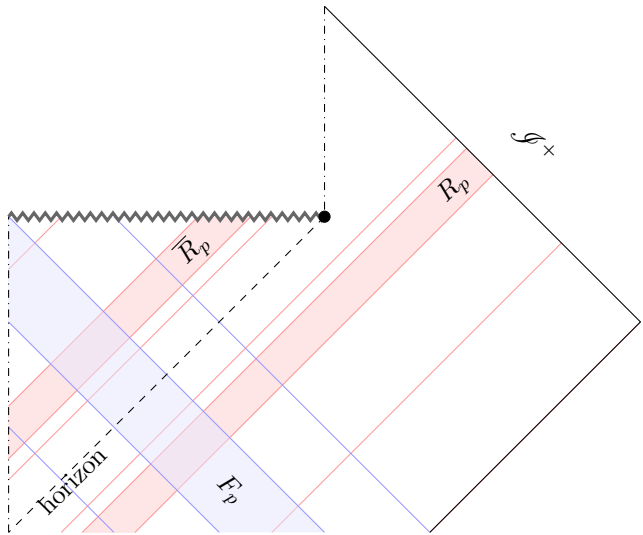

At the semi-classical level, the subsets of Hawking radiation $R_p$ and their partners behind the horizon $\overline R_p$ are illustrated in the Penrose diagram figure \ref{fig3}. 

\subsection{The refined model}

In this section we refine our model of black hole evaporation to take account of energy conservation and the thermal nature of the Hawking radiation. We will work in the adiabatic, or quasi-static, regime where the black hole is evaporating slowly enough that it makes sense to ascribe a slowly varying temperature $T(t)$ to the Hawking radiation determined by the thermodynamic equation of the black hole 
\EQ{
\frac1{T}=\frac{dS_\text{BH}}{dM}\ ,
\label{rib3}
}
where $M$ is the black hole mass\footnote{For a Schwarzschild black hole, $M$ is the mass, while for the charged black hole and the black hole in JT gravity, $M$ is the mass minus the mass of the extremal black hole.}.
The adiabatic regime is where Hawking's calculation derivation is valid. It is defined by the requirement that
\EQ{
S_\text{BH}\gg c\ ,
}
the number of massless fields. In addition, for a semi-classical limit $c\gg1$. 

The time dependence of the energy is determined by the energy flux of the Hawking radiation. Since most of the energy loss occurs in the $s$-wave modes we have effectively a $1+1$-dimensional relativistic gas. We also ignore the possibility for back-scattering of modes and so take a trivial greybody factor. The energy balance equation is then
\EQ{
\frac{dM}{dt}=-\frac{\pi c T^2}{12}
\label{ikl}
}
and given \eqref{rib3} all that it needed to determine the time evolution of $M$, $T$ and $S_\text{BH}$ is the energy dependence of the BH entropy which depends on the nature of the black hole. For example, for Schwarzschild $S_\text{BH}=4\pi GM^2$.

We will model the evaporation in terms as a series of time steps whose size are of the order of the scrambling time of the black hole,
\EQ{
\Delta t\thicksim\frac1{T}\log\frac{S_\text{BH}}c\ .
}
Note that this is time dependent, so the size of the time steps adapt as the evaporation proceeds. 

At each time step, the radiation carries away a small amount of energy in a distribution that is strongly peaked around an average. Therefore, we can model the state of the black hole at each time step as lying in a Hilbert space ${\cal H}_{B_p}$ describing a system with energy in a small window $\Theta_p=[M_p,M_p+\delta M]$. Implicitly, $M_p$ includes the energy of the infalling system $F_p$. In other words, the black hole is in a microcanonical state. The size of the window $\delta M$ is assumed to be small but for simplicity we will assume that it is much larger than the spread of the energy carried away by the radiation at each time step. The fact that the BH entropy is so large means that $\Theta_p$ contains a vast number of states that forms a quasi-continuum. The dimension of this space is exponential in the Bekenstein-Hawking entropy 
\EQ{
d_{B_p}=\frac{C\delta M}{M_p}e^{S_\text{BH}(M_p)}\thicksim e^{S_\text{BH}(M_p)}\ .
}
In the above, $C$ is some constant which we do not have to specify since $S_\text{BH}(M)$ is very large.

The picture of the black hole evolving through a sequence of microcanonical states is of course an approximation which is justified because the radiation emitted during a time step has a sharply defined average energy and a spread that is assumed to be much smaller than the width of the windows $\delta M$. Let us justify this claim. Since the time step, the scrambling time $\Delta t$, is much greater than the thermal scale $T^{-1}$, the energy and entropy of the Hawking radiation follow from the standard statistical mechanics of a relativistic bosonic or fermionic gas (summarized in appendix \ref{a2}). For a bosonic gas
\EQ{
{\cal E}=c{\mathscr V}\int\frac{d\omega}{2\pi}\,\frac{\omega}{e^{\omega/T}-1}=\frac{\pi c\mathscr VT^2}{12}
\label{ted}
}
and the entropy $S_\text{rad}=\pi c{\mathscr V}T/6$, where we identify the volume with the space filled by the gas in the scrambling time, i.e.~$\mathscr V=\Delta t$. In particular, the entropy 
\EQ{
S_\text{rad}=\frac{\pi c\Delta t T}6\thicksim c\log\frac{S_\text{BH}}c\gg1\ .
}
Hence, the Hawking modes emitted in a time step have a large entropy and so can be described thermodynamically. Indeed, the normalized spread of the energy 
\EQ{
\frac{\Delta{\cal E}}{\cal E}\thicksim\frac1{\sqrt{S_\text{rad}}}\ll1\ .
\label{hut}
}
On the other hand, the radiation is a much smaller system than the black hole because
\EQ{
S_\text{BH}\gg c\log\frac{S_\text{BH}}c\ .
}
We will then assume that this spread is much smaller than the microcanonical energy window $\delta M\gg \Delta{\cal E}$ justifying the evaporation as a sequence of microcanonical states. 

The semi-classical state is now a thermofield double with a slowly varying temperature. Taking the basis states $\ket{j_p}$ to be approximate energy eigenstates with eigenvalues ${\cal E}_{j_p}$, we have 
\EQ{
\lambda_J=\frac{e^{-\sum_p{\cal E}_{j_p}/2T_p}}{\sqrt{\cal Z}}\ ,
\label{lambdaJ ref def}
}
where ${\cal Z}=\sum_Je^{-\sum_p{\cal E}_{j_p}/T_p}$ is the partition function which provides normalization. The temperature $T_p$ is the instantaneous temperature of the Hawking radiation given in \eqref{rib3} evaluated at $E=M_p$. The states $\ket{j_p}$ are to be thought of as localized in an outgoing shell of thickness $\Delta t_p$. This is justified because the modes have characteristic momentum $T_p$ and so can be localized on scales $T_p^{-1}$ which is much smaller than $\Delta t_p$.

\subsection{The average state}

Black holes are famously fast scramblers so that over the scrambling time $U_p$ is essentially a random unitary. The question of how random time evolution of a black hole is an interesting question but one can make the hypothesis that for certain quantities it is effectively indistinguishable from a Haar random unitary. In this section, we make that assumption and compute the average of the microscopic state. We note that this was analysed in the basic model in \cite{Akers:2022qdl} and we review the calculation here to set up some notation.

We will need to average quantities over an ${\cal N}\times{\cal N}$ unitary for which the basic results is the integral
\EQ{
\int dU\ U^*_{AB}U_{A'B'}=\frac1{{\cal N}}\delta_{AA'}\delta_{BB'}\ .
\label{in1}
}
We will also need the generalization of this involving $n$ replicas:
\EQ{
\int dU\ \prod_{j=1}^nU^*_{A_jB_j}U_{A'_jB'_j}=\sum_{\sigma,\tau\in S_n}
\prod_{j=1}^n\delta_{A_jA_{\sigma(j)}}\delta_{B_jB_{\tau(j)}}W\hspace{-1mm}g(\sigma\tau^{-1},{\cal N})\ ,
}
where $W\hspace{-1mm}g$ is the Weingarten function \cite{collins:2002, collins:2004}. Note how the integrals over the replicas involves a sum over the elements of the symmetric group $\sigma,\tau\in S_n$ that permute the replicas. We will only need the behaviour in the limit that ${\cal N}$ is large, which picks out the terms with $\sigma=\tau$ for which $W\hspace{-1mm}g(1,{\cal N})=1/{\cal N}$,
\EQ{
\int dU\ \prod_{j=1}^nU^*_{A_jB_j}U_{A'_jB'_j}=\frac1{{\cal N}^n}\sum_{\tau\in S_n}
\prod_{j=1}^n\delta_{A_jA'_{\tau(j)}}\delta_{B_jB'_{\tau(j)}}+\cdots\ ,
\label{in2}
}

Let us consider the microscopic state of the radiation $\rho_R$ and compute its average over the unitaries $U_p$, $p=1,2,\ldots,N$. The ket $\ket{\Psi}$ contributes a $U_p$ and bra $\bra{\Psi}$ a $U_p^\dagger$. The average over $U_p$ then knits together the bra and ket.

Let us focus on the average over $U_p$ of its adjoint action on a operator $f$. Using  \eqref{in1}, we can write this average as
\EQ{
\int dU_p\,U_pfU_p^\dagger=\tr(f)\rho^\text{(mm)}_{R_pB_p}\ .
\label{bob}
}
Here, $\rho^\text{(mm)}_{R_pB_p}$ is the maximally-mixed state on ${\cal H}_{R_p}\otimes{\cal H}_{B_p}$ which in the basic model is,
\EQ{
\rho^{\text{(mm)}}_{R_pB_p}=\frac{\bf 1}{d_{R_p}d_{B_p}}\ .
\label{txt}
}
In the refined model it is the maximally-mixed state in the energy window $\Theta_{p-1}$ embedded in ${\cal H}_{R_p}\otimes{\cal H}_{B_p}$ in such a way as to conserve energy,
\EQ{
\rho^{\text{(mm)}}_{R_pB_p}\ \propto\ \Pi_{\Theta_{p-1}}\ .
\label{rte}
}
where $\Pi_{\Theta_{p-1}}$ is the projector onto the energy window. The following $U_{p+1}$ average then imposes a trace over $B_p$. In the basic model, that gives
\EQ{
\tr_{B_p}\big(\rho^\text{(mm)}_{R_pB_p}\big)=\frac1{d_{R_p}}\sum_{j_p}\ket{j_p}\bra{j_p}\ ,
\label{opa}
}
In the refined model, let us denote a basis of energy eigenstates of $R_p$ as $\ket{j_p}$ with energies ${\cal E}_{j_p}$, then 
\EQ{
\tr_{B_p}\big(\rho^{\text{(mm)}}_{R_pB_p}\big)=e^{-S_\text{BH}(M_{p-1})}\sum_{j_p}e^{S_\text{BH}(M_{p-1}-{\cal E}_{j_p})}\ket{j_p}\bra{j_p}\ .
\label{pes}
}
Implicitly, the sum here is constrained to have $M_{p-1}-{\cal E}_{j_p}\in{\Theta}_p$. We can now follow the standard route for deriving the canonical ensemble of a small subsystem of a larger system in a microcanonical state \cite{Goldstein:2005aib}, in our case the maximally mixed state. Since the radiation subsystem is much smaller then the black hole, we can expand $S_\text{BH}(M_{p-1}-{\cal E}_{j_p})\approx S_\text{BH}(M_{p-1})-{\cal E}_{j_p}/T_p$ where the temperature is defined in the standard way via the thermodynamic equation \eqref{rib3} for a black hole of mass $M_{p-1}$. Then we can extend the restricted sum over ${\cal E}_{j_p}$ to be unrestricted because terms for which  $M_{p-1}-{\cal E}_{j_p}\not\in\Theta_p$ are heavily suppressed. This gives the familiar approximation, namely the canonical state
\EQ{
\tr_{B_p}\big(\rho^{\text{(mm)}}_{R_pB_p}\big)\approx\sum_{j_p}\frac{e^{-{\cal E}_{j_p}/T_p}}{{\cal Z}_p}\ket{j_p}\bra{j_p}\ ,
\label{opb}
}
where ${\cal Z}_p=\sum_{j_p}e^{-{\cal E}_{j_p}/T_p}$.

If we now assemble the expressions \eqref{opa} and \eqref{opb} for all the time steps, to find the average state of the radiation
\EQ{
        \overline{\rho_R}&=\frac1{d_R}\sum_J\ket{J}\bra{J} \qquad \qquad \qquad \, \text{(basic)}\ , \\[5pt]
        \overline{\rho_R}&=\sum_J\frac{e^{-\sum_p{\cal E}_{j_p}/T_p}}{\cal Z}\ket{J}\bra{J} \quad \qquad \text{(refined)}\ .
\label{tot}
}
Hence the averaged microscopic state $\rho_R$ is precisely the semi-classical state $\rho_R^\text{sc}$ as stated in \eqref{tot2}. Deviations from the average arise because of the non-isometric nature of the map. These have been analysed in the basic model in \cite{Akers:2022qdl}.

\section{Entropies}\label{s4}

We can calculate the entropy of the microscopic state reduced on any subset 
\EQ{
A\subset \{R_1,\ldots,R_N,B\}\ .
} 
The strategy is to first calculate the R\'enyi entropies which can be defined by introducing $n$ replicas of the Hilbert space
\EQ{
e^{(1-n)S^{(n)}(A)}=\tr(\rho_A^n)=\tr^{(n)}\sigma_1^{[R_1]}\cdots\sigma_N^{[R_N]}\tau_{N+1}^{[B]}\ \ket{\Psi}\bra{\Psi}^{\otimes n}\ ,
\label{put}
}
where the $\sigma_p$ and $\tau_{N+1}$ are elements of the symmetric group $S_n$. The superscripts, e.g.~$\sigma_p^{[R_p]}$, on these elements indicate which subspace of the replicated Hilbert space the element acts on where it is ambiguous. These elements are taken to be either the identity element $e$ or the cyclic permutation $\eta$ according to the definition of the subset $A$
\EQ{
A=\big\{R_p\ \big|\ \sigma_p=\eta\ ,\ p=1,2,\ldots,N\big\}\cup\big\{B\ \big|\ \tau_{N+1}=\eta\big\}\ .
\label{boo}
}

The R\'enyi entropies are known to be self-averaging in the ensemble of the unitaries $U_p$ (e.g.~\cite{Liu:2020jsv}) and so we will calculate the ensemble average of \eqref{boo} and take this to describe a typical element of the ensemble. The integrals we need are given in \eqref{in2} which capture the leading order behaviour when the Hilbert spaces have a large dimension. 

Using \eqref{in2}, the average over the unitary $U_p$ acting in a replicated Hilbert space at large $d_{R_p}d_{B_p}$ of adjoint action is given by a sum over elements of the symmetric group $S_n$, 
\EQ{
\int dU_p\,U_p^{\dagger\,\otimes n} \,f\, U_p^{\otimes n}  = \sum_{\tau_p\in S_n}\Big\{ \tr^{(n)}\tau_p^{[B_{p-1}F_{p-1}]} f\Big\}\,(\tau_p^{[R_pB_p]})^{-1}\ \rho_{R_pB_p}^{\text{(mm)}\ \otimes n}+\cdots\ ,
\label{pop2}
}
for some $f$ in the replicated Hilbert space. So each average over $U_p$ comes with a sum over an element of the symmetric group $\tau_p\in S_n$. In the above, $\rho_{R_pB_p}^\text{(mm)}$ is the maximally mixed state of ${\cal H}_{R_p}\otimes{\cal H}_{B_p}$ as in \eqref{txt}, while for the refined model, it is the subspace with energy in the window $\Theta_{p-1}$ as in \eqref{rte}. The ellipsis stand for subleading corrections, suppressed by inverse powers of $d_{B_{p-1}}d_{F_{p-1}}$, that we will not keep track of in our analysis. $\tr^{(n)}$ is the trace defined on the replicated Hilbert space.

Applying \eqref{pop2} for all $p$, it becomes apparent that the average of \eqref{put} breaks up into a set of building blocks:
\EQ{
\overline{e^{(1-n)S^{(n)}(A)}}=\sum_{\tau_1,\ldots,\tau_N\in S_n}{\cal Z}_1\cdots{\cal Z}_N\ ,
\label{ruc}
}
where
\EQ{
{\cal Z}_p=\tr^{(n)}\sigma_p^{[R_p]}\tau_{p+1}^{[B_pF_p]}(\tau_p^{[R_pB_p]})^{-1}\big(\rho^\text{(mm)}_{R_pB_p}\otimes\rho^\text{sc}_{F_p}\big)^{\otimes n}\ ,
\label{yet}
}
where $\rho^\text{sc}_{F_p}=\ket{s_p}\bra{s_p}$ is the semi-classical state of the infalling system $F_p$. In the last step $p=N$ this piece is missing, there is no $F_N$. The traces over ${\cal H}_{F_p}$ are trivial because the states $\rho^\text{sc}_{F_p}$ are pure and so $\tr^{(n)}(\sigma\rho_{F_p}^{\text{sc}\,\otimes n})=1$, for any $\sigma\in S_n$. This includes the initial state in ${\cal H}_{F_0}$ that collapsed to form the black hole. Hence, the building block \eqref{yet} can be written more simply as
\EQ{
{\cal Z}_p=\tr^{(n)}\sigma_p^{[R_p]}\tau_{p+1}^{[B_p]}(\tau_p^{[R_pB_p]})^{-1}\rho^{\text{(mm)}\ \otimes n}_{R_pB_p}\ .
\label{yet2}
}
\begin{figure}[h]
\begin{center}
\begin{tikzpicture} [scale=0.7]
\draw[fill = Plum!10!white,line width=0.6mm,rounded corners = 0.2cm] (-1,-1) rectangle (1,1);
\draw[-triangle 60,line width=0.4mm] (-3.5,0) -- (-1,0);
\draw[decorate,line width=0.4mm,decoration={snake,amplitude=0.03cm},-triangle 60] (1,0.5) -- (4.5,2.2) node[right,sloped] {${\cal H}_{R_{p}}$};
\draw[-triangle 60,line width=0.4mm] (1,-0.5)  -- (5,-0.5) node[sloped,midway,below] {${\cal H}_{B_{p}}$};
\node at (0,0) {$U_p$};
\draw[-triangle 60,line width=0.4mm] (-3.5,-1.5)  -- (-1,-0.5);
\begin{scope}[xshift=6cm,yshift=-1cm]
\draw[fill = Plum!10!white,line width=0.6mm,rounded corners = 0.2cm] (-1,-1) rectangle (1,1);
\draw[decorate,line width=0.4mm,decoration={snake,amplitude=0.03cm},-triangle 60] (1,0.5) -- (3.5,1.5);
\draw[-triangle 60,line width=0.4mm] (1,-0.5)  -- (4,-0.5);
\node at (0,0) {$U_{p+1}$};
\draw[-triangle 60,line width=0.4mm] (-4.5,-2)  -- (-1,-0.5) node[sloped,midway,below] {${\cal H}_{F_p}$};
\node[rotate=30] at (-1.9,-0.48) {$\tau_{p+1}$};
\node[blue,left] at (-4.5,-2.2) {$\rho_{F_p}=\ket{s_p}\bra{s_p}$};
\end{scope}
\node at (2,0) {$\tau_p^{-1}$};
\node[rotate=30] at (1.9,1.5) {$\tau_p^{-1}$};
\node[rotate=30] at (3.4,2.1) {$\sigma_p$};
\node at (4,-0.15) {$\tau_{p+1}$};
\draw[very thick,blue,dotted] (1.2,-2) -- (1.2,2) node[above] {$\rho_{R_pB_p}^\text{(mm)}$};
\end{tikzpicture}
\caption{\footnotesize Assembling the ingredients for the building block in \eqref{yet}.} 
\label{fig2}
\end{center}
\end{figure}
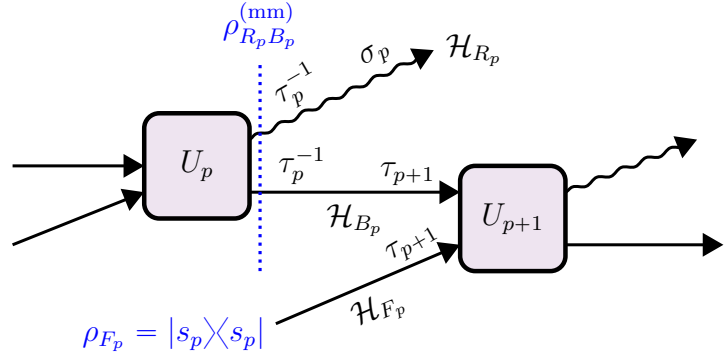
The expression for the building block ${\cal Z}_p$ can also be interpreted in terms of the equilibration ansatz of \cite{Liu:2020jsv} as an alternative to the unitary averages. In this interpretation, the pure state of the black hole at time $t_{p-1}$ equilibrates over the next time step meaning that for certain observables it is indistinguishable from an equilibrium state, in this case precisely the maximally mixed state $\rho_{R_pB_p}^\text{(mm)}$ \eqref{txt}, or \eqref{rte} in the refined model.

In the basic model, it is then straightforward to evaluate the building block \eqref{yet2},
\EQ{
{\cal Z}_p=\exp\Big[(k(\tau_{p+1}\tau_p^{-1})-n)S_\text{BH}(M_p)+(k(\sigma_p\tau_p^{-1})-n)S_\text{rad}(R_p)\Big]\ ,
\label{rib2}
}
where $k(\sigma)$ is the number cycles of the element $\sigma$ and with $S_\text{BH}(M_p)=\log d_{B_p}$ and $S_\text{rad}(R_p)=\log d_{R_p}$. Then plugging into \eqref{ruc} gives the final result
\EQ{
\overline{e^{(1-n)S^{(n)}(A)}}=\sum_{\tau_1,\ldots,\tau_N\in S_n}\ e^{(1-n)S^{(n)}_{\{\tau_p\}}(A)}\ ,
\label{jug}
}
where we have defined
\EQ{
S_{\{\tau_p\}}^{(n)}(A)=\frac1{n-1}\sum_{p=1}^N\Big\{d(\tau_{p+1},\tau_p)S_\text{BH}(M_p)+d(\sigma_p,\tau_p)S_\text{rad}(R_p)\Big\}\ ,
\label{jut}
}
where $d(\sigma,\pi)=n-k(\sigma\pi^{-1})$ is the Cayley distance between elements of $S_n$.\footnote{Alternatively, the Cayley distance $d(\sigma, \pi)$ may be defined as the minimal number of transpositions required to go between $\sigma$ and $\pi$.}

\subsection{Refined model}

The refined model is rather more complicated because of the need to enforce energy conservation. The R\'enyi entropies now involve a sum over both the energies ${\cal E}_{j_p}$ and the elements of the symmetric group $\tau_p$,
\EQ{
\overline{e^{(1-n)S^{(n)}(A)}}=\sum_{\{j_p\}}\,\sum_{\{\tau_p\}\in S_n}\,\prod_{p=1}^n{\cal Z}_p\ =\sum_{\{j_p\}}\sum_{\{\tau_p\}\in S_n}\ e^{(1-n)S^{(n)}_{\{\tau_p\}}(A)}
\label{mim}
}
where the building block is
\EQ{
{\cal Z}_p=\frac{d_{R_{p}}({\cal E}_{j_p})^{k(\sigma_p\tau_p^{-1})}d_B(M_{p-1}+E_p-{\cal E}_{j_p})^{k(\tau_{p+1}\tau_p^{-1})}}
{\big(\sum_{j_p}\ d_{R_{p}}({\cal E}_{j_p})d_B(M_{p-1}+E_p-{\cal E}_{j_p})\big)^n}\ .
\label{pop}
}
where $E_p$ is the energy of the infalling system $F_p$. Note that the mass of the black hole depends implicitly on the energy of the radiation emitted up to that point
\EQ{
M_p=M_0+\sum_{q=1}^p(E_q-{\cal E}_{j_q})\ ,
\label{econ}
}
a point that must be born in mind when we perform the saddle point approximation.

The denominator in \eqref{pop} can be evaluated by a saddle point approximation where the sum is replaced by an integral over a continuous variable ${\cal E}_p$. In particular, the radiation can be described thermodynamically in the way summarized in appendix \ref{a2} and the entropy
\EQ{
\log d_{R_{p}}({\cal E})=2\sqrt{\mu_{p}{\cal E}_p}\qquad\text{where}\qquad \mu_{p}=\frac{\pi c\Delta t_{p}}{12}\ .
\label{fit}
}
Since the saddle point value of the energy is much smaller than the black hole mass, the saddle point equation is 
\EQ{
\sqrt{\frac{\mu_p}{{\cal E}_p}}=-\frac{dS_\text{BH}(M_{p-1}+E_p-{\cal E}_p)}{d{\cal E}_p}\approx\frac1{T_p}\qquad\implies\qquad {\cal E}_p=\mu_pT_p^2\ ,
}
where $T_p$ defined in \eqref{rib3} is precisely the temperature of the Hawking radiation $R_p$. The average energy of the radiation emitted ${\cal E}_p$ and the infalling energy $E_p$ are assumed to be much smaller than the black hole mass . Hence, we have
\EQ{
\sum_{j_p}\ d_{R_{p}}({\cal E}_{j_p})d_B(M_{p-1}+E_p-{\cal E}_{j_p})\approx d_B(M_{p-1})e^{S_\text{rad}(R_p)/2+E_p/T_p}\ ,
\label{jes}
}
where the saddle point value of the entropy is 
\EQ{
S_\text{rad}(R_p)=2\mu_pT_p=\frac{\pi c\Delta t_pT_p}6\ .
}
This and ${\cal E}_p$ above are the familiar expressions for the entropy and energy of a volume $\Delta t_p$ of a relativistic gas in $1+1$ dimensions in a volume $\mathscr V=\Delta t_p$ (as reviewed in appendix \ref{a2}). The saddle point approximation is, of course, just the conventional way of deriving the Legendre transformation between the internal energy and free energy in thermodynamics and is justified precisely because the spread in the energy is small \eqref{hut}. 

For later use, note that 
\EQ{
d_B(M_p)=d_B(M_{p-1}+E_p-{\cal E}_p)\approx d_B(M_{p-1}+E_p)e^{-S_\text{rad}(R_p)/2}
}
and so
\EQ{
S_\text{BH}(M_{p-1}+E_p)-S_\text{BH}(M_p)=\frac{S_\text{rad}(R_p)}2\ ,
\label{pux}
}
which is the familiar relation for a model of black hole evaporation in the $s$-wave approximation and with no back scattering (i.e.~grey body factor). Note that it implies that the evaporation is irreversible.
 
We now proceed to evaluate the sums of the energies in \eqref{mim} by similar saddle point approximations. 
After we replace the sums by integrals over ${\cal E}_p$, the exponent of the integrand is
\EQ{
(1-n)S_{\{\tau_p\}}^{(n)}(A)&=\sum_{p=1}^N\Big\{2(n-d(\sigma_p,\tau_p))\sqrt{\mu_p{\cal E}_p}-d(\tau_{p+1},\tau_p)S_\text{BH}(M_0)\\[5pt] &-\Big(n-\sum_{q=p}^Nd(\tau_{q+1},\tau_q)\Big)\frac{{\cal E}_p}{T_p}-\sum_{q=p}^Nd(\tau_{q+1},\tau_q)\frac{E_p}{T_p}-\frac n2S_\text{rad}(R_p)\Big\}\ .
}
It is now simple to compute the saddle point equations for the energies ${\cal E}_p$. In the regime of slow evaporation we can ignore the ${\cal E}_p$ dependence of the temperatures $T_p$. The saddle point values are found to be
\EQ{
{\cal E}_p=\mu_pT_p^2\Big(\frac{n-d(\sigma_p,\tau_p)}{n-\sum_{q=p}^N d(\tau_{q+1},\tau_q)}\Big)^2\ ,
}
where for consistency the saddles must have
\EQ{
n>\sum_{q=1}^N d(\tau_{q+1},\tau_q)\ .
}
The contribution of this saddle to the R\'enyi entropy is
\EQ{
S^{(n)}_{\{\tau_p\}}(A)&=\frac1{n-1}\sum_{p=1}^N\Big\{d(\tau_{p+1},\tau_p)S_\text{BH}(M_0) +\sum_{q=p}^Nd(\tau_{q+1},\tau_q)\frac{E_p}{T_p}\\[5pt] &\qquad\qquad
+\frac12\Big(n-\frac{(n-d(\sigma_p,\tau_p))^2}{n-\sum_{q=p}^N d(\tau_{q+1},\tau_q)}\Big)S_\text{rad}(R_p)
\Big\}\ .
}
We can re-write this by noting that \eqref{pux} implies
\EQ{
S_\text{BH}(M_p)=S_\text{BH}(M_0)+\sum_{q=1}^p\Big(\frac{E_q}{T_q}-\frac{S_\text{rad}(R_q)}2\Big)\ ,
}
as
\EQ{
S^{(n)}_{\{\tau_p\}}(A)&=\frac1{n-1}\sum_{p=1}^N\Big\{d(\tau_{p+1},\tau_p)S_\text{BH}(M_p) \\[5pt] &+\frac{2nd(\sigma_p,\tau_p)-d(\sigma_p,\tau_p)^2-\big(\sum_{q=p}^Nd(\tau_{q+1},\tau_q)\big)^2}{2\big(n-\sum_{q=p}^N d(\tau_{q+1},\tau_q)\big)}S_\text{rad}(R_p)\Big\}\ ,
\label{wil}
}
which is the refined model generalization of \eqref{jut}.

\subsection{Relation to the island formalism}

We interpret \eqref{jug} as being a sum over saddles of the (Lorentzian) gravitational path integral in the semi-classical limit, labelled by the elements $\{\tau_p\}$. In this limit, the entropies $S_\text{BH}(M_p)$ and $S_\text{rad}(R_p)$ are very large. If we avoid the crossover regimes when saddles are degenerate, it turns out that only a much smaller number of terms can actually dominate in the sum, namely, those for which each $\tau_p$, $p=1,\ldots,N$, is equal to $e$ or $\eta$ only, the identity and cyclic permutations, respectively. This is proved in appendix \ref{a1} for the basic model. The $\{e,\eta\}$ dominance means that the saddles that dominate respect the ${\mathbb Z}_n$ cyclic symmetry of the replicas mirroring the symmetry of the replica wormholes of \cite{Penington:2019kki,Almheiri:2019qdq}, or, equivalently, we can interpret the average over unitaries to be equivalent to the average over baby universe states (see \cite{Marolf:2020rpm,Maxfield:2022sio}).

For the refined model, the discussion is very similar. Indeed each element in the energy sum in \eqref{mim} behaves like a basic model, and therefore we can again invoke the fact that $\tau_p$ is dominated by $\tau_p\in\{e,\eta\}$, which will be valid as long we are not in the vicinity of a crossover of saddles.\footnote{Notice that we don't risk of having a crossover at every time step since we assumed that each energy window is small \eqref{hut}. }

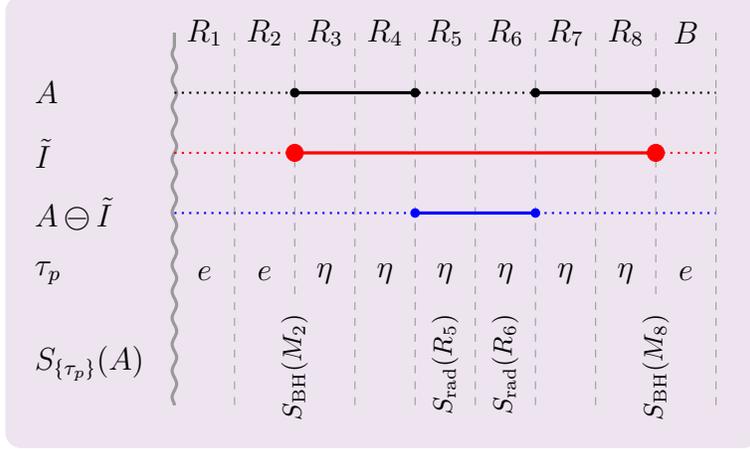
\begin{figure}
\begin{center}
	\begin{tikzpicture} [scale=0.8]
	\filldraw[fill = Plum!10!white, draw = Plum!10!white, rounded corners = 0.2cm] (-2.8,1.6) rectangle (9.7,-5.9);
	\draw[decorate,very thick,black!40,decoration={snake,amplitude=0.03cm}] (0,-5.2) -- (0,1);
	\draw[dotted,thick] (0,0) -- (9,0);
	\draw[dotted,thick,red] (0,-1) -- (9,-1);
	\draw[dotted,thick,blue] (0,-2) -- (9,-2);
	\draw[very thick] (6,0) -- (8,0);
	\draw[very thick] (2,0) -- (4,0);	
	\draw[very thick,red] (2,-1) -- (8,-1);
\draw[black!40,dashed] (1,1) -- (1,-5.2);
\draw[black!40,dashed] (2,1) -- (2,-3.4);
\draw[black!40,dashed] (3,1) -- (3,-5.2);
\draw[black!40,dashed] (4,1) -- (4,-5.2);
\draw[black!40,dashed] (5,1) -- (5,-5.2);
\draw[black!40,dashed] (6,1) -- (6,-5.2);
\draw[black!40,dashed] (7,1) -- (7,-5.2);
\draw[black!40,dashed] (8,1) -- (8,-3.4);
\draw[black!40,dashed] (9,1) -- (9,-5.2);			
	\filldraw[black] (6,0) circle (2pt);
	\filldraw[black] (8,0) circle (2pt);
	\filldraw[black] (2,0) circle (2pt);
	\filldraw[black] (4,0) circle (2pt);	
	\filldraw[red] (2,-1) circle (4pt);
	\filldraw[red] (8,-1) circle (4pt);	
	\draw[very thick,blue] (4,-2) -- (6,-2);
	\filldraw[blue] (4,-2) circle (2pt);
	\filldraw[blue] (6,-2) circle (2pt);
	\node[right] at (-2.5,0) {$A$};
	\node[right] at (-2.5,-1) {$\tilde I$};
	\node[right] at (-2.5,-2) {$A\ominus\tilde I$};
	\node[right] at (-2.5,-3) {$\tau_p$};	
	\node[right] at (-2.5,-4.5) {$S_{\{\tau_p\}}(A)$};	
\node at (0.5,1) {$R_1$};
\node at (1.5,1) {$R_2$};
\node at (2.5,1) {$R_3$};
\node at (3.5,1) {$R_4$};
\node at (4.5,1) {$R_5$};
\node at (5.5,1) {$R_6$};
\node at (6.5,1) {$R_7$};
\node at (7.5,1) {$R_8$};
\node at (8.5,1) {$B$};
\node at (0.5,-3) {$e$};
\node at (1.5,-3) {$e$};
\node at (2.5,-3) {$\eta$};
\node[rotate=90,left] at (2,-3.5) {\footnotesize$S_\text{BH}(M_2)$};
\node at (3.5,-3) {$\eta$};
\node at (4.5,-3) {$\eta$};
%
\node at (5.5,-3) {$\eta$};
\node[rotate=90,left]  at (4.5,-3.5) {\footnotesize$S_\text{rad}(R_5)$};
\node at (6.5,-3) {$\eta$};
\node[rotate=90,left]  at (5.5,-3.5) {\footnotesize$S_\text{rad}(R_6)$};
\node at (7.5,-3) {$\eta$};
\node at (8.5,-3) {$e$};
\node[rotate=90,left]  at (8,-3.5) {\footnotesize $S_\text{BH}(M_8)$};
	%
	%
	\end{tikzpicture}
\caption{\footnotesize An example of a saddle for the model with $N=8$ time steps, with some choice of the set $A=R_3\cup R_4\cup R_7\cup R_8$, as shown, with an island-in-the-stream $\tilde I$. Note that $\partial\tilde I\subset\partial A$. The contributions to the entropy from each time step are shown and summing these up gives $S_I(A)=S_\text{BH}(M_2)+S_\text{BH}(M_8)+S_\text{rad}(R_5\cup R_6)$. Note that the last term is $S_\text{rad}(A\ominus\tilde I)$.
}
\label{fig2}
\end{center}
\end{figure}

The expression for the von Neumann entropy of our chosen subset $A\subset R\cup B$ is obtained from \eqref{jut} and \eqref{wil}  in the limit $S(A)=\lim_{n\to1} S^{(n)}(A)$ and has the form of a minimization problem over the $2^N$ choices  $\tau_p\in\{e,\eta\}$. Indeed notice that when $\sigma,\pi\in\{e,\eta\}$, we can write
\EQ{
d(\sigma,\pi)=(n-1)(1-\delta_{\sigma\pi})\ ,
\label{btq}
}
which facilitates the evaluation of the Cayley distances in the $n\to1$ limit of \eqref{jut} and  \eqref{wil}. In both models, the von Neumann entropy is given by
\EQ{
 S(A)= \min_{\{\tau_p\}}S_{\{\tau_p\}}(A)=\min_{\{\tau_p\}} \Big\{\sum_{p=1}^{N}(1-\delta_{\tau_{p+1}\tau_p})S_\text{BH}(M_p)+(1-\delta_{\sigma_p\tau_p})S_\text{rad}(R_p)\Big\} .
\label{tuz}
} 

The resemblance of this equation to the QES formula described in the introduction for a slowly evaporating black hole \eqref{twm} becomes more apparent if we set
\EQ{
 S_I(A) \equiv S_{\{\tau_p\}}(A) .
\label{xin}
}
where $I$ is defined in both models as
\EQ{
I=\bigcup_{p\in\Phi}\big(\overline R_p\cup F_{p-1}\big)\ .
\label{bom}
}
with $\Phi=\big\{p\ |\ \tau_p=\eta\big\}$.  The $I$ that minimizes \eqref{xin} is called the `entanglement island' or `island', for short, will be denoted $I(A)$.
Even if in principle we have $2^N$ possible saddles, most of them will not contribute since terms with $\tau_p \neq \tau_{p+1}$ are not favourable because of the black hole entropy being big. One can check that the only saddles that are not trivially suppressed are the one where $\tau_p$ changes in correspondence with a change in $\sigma_p$, which is an analog of the condition \eqref{eq:extr}. See figure \ref{fig2} for an example where $\Phi=\{3,4,5,6,7,8\}$.

In order to make more transparent the identification of \eqref{tuz} with the QES formula \eqref{twm} for the $A$ that we have chosen, we can also notice that the second term is a discrete version of the continuum expression $S_\text{rad}(A\ominus \tilde I)$ where we identify the island-in-the-stream as the reflection of the island $I$ in the horizon and then projected onto $\mathscr I^+$, so each $\overline R_p$ gets mapped to $R_p$:
\EQ{
\tilde I=\bigcup_{p\in\Phi}R_p\ .
\label{lip}
}
On the other hand, the first term can be written in terms of the BH entropy at the outgoing EF coordinates of QES $u_{\partial I}$. We can then parametrize the entropy of the black hole with its mass at outgoing time $u$. Notice also that the infalling states in \eqref{bom} are shifted by $p\to p-1$. This is how the model accounts for the fact that infalling coordinate $v$ of the QES are shifted relative to the outgoing coordinate $u$ by the scrambling time \eqref{vic}, precisely the size of the time steps in the model.

In the next sections, we will enforce our definition of the entanglement island \eqref{bom} studying when it is possible to reconstruct an unitary acting on the radiation, which is equivalent to the well known statement that the island is in the entanglement wedge of the radiation.  
Specifically, since the emitted radiation is in both the semi-classical and microscopic descriptions, we include it in its own entanglement wedge
\EQ{
{\cal W}(A)=I(A)\cup(A\cap R)\ .
}
Although we call $I(A)$ the island, strictly speaking, this only applies when subsets of $I(A)$ are separated from the rest of the entanglement wedge by QES.\footnote{For example, when $A=B$, the black hole before the Page time has $I(B)={\cal W}(B)=\overline R\cup F$ which is not an island in the strict sense.} 

\section{Information recovery and reconstruction}\label{s5}

In this section, we consider the fate of an infalling system, Hayden and Preskill's diary for instance \cite{Hayden:2007cs}. A version of this problem was considered for the basic (or BRU) model in section 7 of \cite{Akers:2022qdl}. The purpose of this section is to extend this analysis to reconstruction on subsets of the radiation, which we verify is consistent with the QES formula discussed in section \ref{s4}, and to extend this analysis to the refined model. It is worth noting that we employ a different method, as compared with \cite{Akers:2022qdl}, to show when reconstruction is possible by employing a replica trick to calculate the trace distance to find when certain decoupling conditions hold. Finally, we point out in section \ref{s5.1} how a different approach which uses standard bounds on the trace distance can not always be used in the refined model.

We will focus on a single system that falls in during the $q^\text{th}$ time step. For simplicity, we will avoid the case that this is the last time step, in other words we will take $q<N$. The idea is to consider a family of infalling states $W\ket{s_q}$ for a unitary $W$ and fixed state $\ket{s_q} \in F_q$. This gives a family of microscopic states $\ket{\Psi(W)}$. The physical question is, can the effect of the unitary $W$ be achieved by a local action on the radiation or the black hole? This will inform us as to when the information in $F_q$ has been teleported out of the black hole. More specifically, when can the action of $W$ be {\it reconstructed\/} on $A=R$ or $B$, or a subset thereof, in the sense that there exists a unitary $W_A$ acting on $A$ such that
\EQ{
W_A\ket{\Psi}\overset?=\ket{\Psi(W)}\ .
\label{xea}
}
This is the state-specific notion of reconstruction described in \cite{Akers:2022qdl}. The above implies that $W_A$ acts on the reduced state on $A$ via the adjoint action
\EQ{
\rho_A(W)=W_A\rho_AW_A^\dagger\ ,
\label{xeb}
}
while the reduced state on the complement $\overline A$ is invariant
\EQ{
\rho_{\overline A}(W)=\rho_{\overline A}\ .
\label{xec}
}
In fact this {\it decoupling condition\/} on $\overline A$ implies the existence of $W_A$ in \eqref{xea}. This can be seen using the Schmidt decomposition. The decoupling condition implies that if $\ket{\Psi}=\sum_j\sqrt{p_j}\ket{j}_A\ket{j}_{\overline A}$ then $\ket{\Psi(W)}=\sum_j\sqrt{p_j}\ket{j}'_A\ket{j}_{\overline A}$. It follows that $W_A=\sum_j\ket{j}'_A\bra{j}$ acting on the subspace of ${\cal H}_A$ spanned by the Schmidt states $\ket{j}_A$ although it can be extended to a unitary acting on ${\cal H}_A$. Acting within the subspace, we can write explicitly,
\EQ{
W_A=\tr_{\overline A}\,\ket{\Psi(W)}\bra{\Psi}\rho_{\overline A}^{-1}\ .
}
The Schmidt basis states depend implicitly on the infalling state $\ket{s_q}$ and so the construction of $W$ is `state dependent' in this sense.  It is an interesting question if the construction can be extended to any operator acting on any state of the infalling system in ${\cal H}_{F_q}$ and thereby be state independent, at least in this limited sense. In fact, the construction above can be seen as a special case of the Petz map and, indeed, there is a more general state-independent construction \cite{us}.

We cannot expect the conditions \eqref{xea} and \eqref{xec} to hold exactly and approximate forms of these conditions are formulated in \cite{Akers:2022qdl}. However, we will work to leading order in the semi-classical limit and we will not need these approximate forms in our analysis.

The decoupling condition is therefore key to reconstructing that action of $W$ on either the radiation or the black hole. Hence, we need to calculate the difference between the states $\rho_A(W)$ and $\rho_A$. This can be measured by the trace norm\footnote{For Hermitian operators the trace norm is equal to $\NORM{{\cal O}}_1=\sum_j|\lambda_j|$, where $\lambda_j$ are the eigenvalues of ${\cal O}$.} difference $\NORM{\sigma-\rho}_1$ or the quantum fidelity $f(\sigma,\rho)$. Both are tractable in our models when averaged over the unitary evolution to leading order in the semi-classical limit where they can be computed using the replica method and an analytic continuation. For the trace norm difference, we take an even number of replicas and then take an analytic continuation,
\EQ{
\NORM{\sigma-\rho}_1=\tr\sqrt{(\sigma-\rho)^2}=\lim_{n\to\frac12}\tr^{(2n)}\eta\big(\sigma-\rho\big)^{\otimes 2n}
\label{jg1}
}
and similarly for the quantum fidelity,
\EQ{
f(\sigma,\rho)\equiv \tr\sqrt{\sqrt\rho\sigma\sqrt\rho}=\lim_{n\to\frac12}\tr^{(2n)}\eta\big(\sigma\otimes\rho\big)^{\otimes n}\ .
}
In our context, there is a subtlety in that the analytic continuations must be taken {\it after\/} the semi-classical limit has picked out a dominant saddle otherwise saddles would become degenerate. We should also emphasize that what we are actually calculating are the unitary averages of the replica expressions before taking the limits $n\to\frac12$. This is in the same spirit as calculating the averages the exponents of the R\'enyi entropies as in \eqref{ruc} before taking the limit $n\to1$ to recover the von Neumann entropy. In section \ref{s5.1} we compute an upper bound on the trace norm which does not require the $n\to\frac12$ limit.

Let us compute the average of the trace difference in \eqref{jg1}. The computation is similiar to that of the R\'enyi entropy via $\tr\rho_A^n$. In fact, since $W$ acts locally on ${\cal H}_{F_q}$, only the $q^\text{th}$ time step is modified:
\EQ{
{\cal Z}_q&\longrightarrow\tr^{(2n)}\sigma_q^{[R_q]}\tau_{q+1}^{[B_qF_q]}(\tau_q^{[R_qB_q]})^{-1}\big(\rho^\text{(mm)}_{R_qB_q}\otimes(\rho_{F_q}^\text{sc}(W)-\rho_{F_q}^\text{sc})\big)^{\otimes2n}\\[5pt]
&={\cal Z}_q\,\tr^{(2n)}\tau_{q+1}\big(\rho_{F_q}^{\text{sc}}(W)-\rho_{F_q}^{\text{sc}}\big)^{\otimes2n}\ ,
\label{tum}
}
where we separated out the trace over the replicas of $F_q$ where $W$ acts and the quantity ${\cal Z}_q$ is the original quantity in the entropy calculation defined in \eqref{yet2}. The contribution from the other time steps $p\neq q$ are precisely as for the entropy \eqref{yet2}. Hence, assembling all the pieces gives 
\EQ{
\overline{\NORM{\rho_A(W)-\rho_A}_1}=\lim_{n\to\frac12}\sum_{\tau_1,\ldots,\tau_N\subset\{e,\eta\}}e^{(1-2n)S^{(2n)}_{\{\tau_p\}}(A)}
\,\tr^{(2n)}\tau_{q+1}\big(\rho_{F_q}^{\text{sc}}(W)-\rho_{F_q}^{\text{sc}}\big)^{\otimes2n}\ .
\label{bez}
}
Now we have to be careful to take the semi-classical limit before taking the analytic continuation $n\to\frac12$. The semi-classical limit picks out a dominant term in the sum over the elements $\tau_p$ and, in particular, fixes $\tau_{q+1}$. Hence, 
\EQ{
\boxed{\overline{\NORM{\rho_A(W)-\rho_A}_1}=\lim_{n\to\frac12}\tr^{(2n)}\tau_{q+1}\big(\rho_{F_q}^\text{sc}(W)-\rho_{F_q}^\text{sc}\big)^{\otimes2n}}
\label{epi}
}

One can follow the same steps for the average of the quantum fidelity. Once again the contribution comes entirely from the  $q^\text{th}$ time step which is modified as
\EQ{
{\cal Z}_q&\longrightarrow\tr^{(2n)}\sigma_q^{[R_q]}\tau_{q+1}^{[B_qF_q]}(\tau_q^{[R_qB_q]})^{-1}\big(\rho^\text{(mm)}_{R_qB_q}\big)^{\otimes2n}\otimes\big(\rho_{F_q}^\text{sc}(W)\otimes\rho_{F_q}^{\text{sc}}\big)^{\otimes n}\\[5pt]
&={\cal Z}_q\,\tr^{(2n)}\tau_{q+1}\big(\rho_{F_q}^\text{sc}(W)\otimes\rho_{F_q}^{\text{sc}}\big)^{\otimes n}\ ,
\label{tom}
}
leading to
\EQ{
\boxed{\overline{f(\rho_A(W),\rho_A)}=\lim_{n\to\frac12}\tr^{(2n)} \tau_{q+1}\big(\rho_{F_q}^\text{sc}(W)\otimes\rho_{F_q}^{\text{sc}}\big)^{\otimes n}}
\label{tom2a}
}

Let us now evaluate our results above. When $F_q\not\in{\cal W}(A)$, it follows that the dominant saddle has $\tau_{q+1}=e$. For the trace norm difference \eqref{epi}, this gives an expression that is clearly seen to vanish
\EQ{
\overline{\NORM{\rho_A(W)-\rho_A}_1}&=\lim_{n\to\frac12}\tr^{(2n)}\big(\rho_{F_q}^\text{sc}(W)-\rho_{F_q}^\text{sc}\big)^{\otimes2n}\\[5pt] &=\big|\tr(\rho_{F_q}^\text{sc}(W)-\rho_{F_q}^\text{sc})\big|=0\ .
\label{tum2}
}
This proves the decoupling condition in terms of the trace norm. On the other hand, for the fidelity \eqref{tom2a},\footnote{The fidelity plays an important role in quantum hypothesis testing, which is the task of making a measurement to distinguish between two quantum states given that the actual state is one of them. The fidelity bounds the error on the optimal measurement. We expect that the corrections to \eqref{tom2} are non-perturbatively suppressed in the semi-classical limit, as in \cite{Kudler-Flam:2021alo, Kudler-Flam:2022irq}. If so, this would imply that whilst it is not possible to distinguish the two states given a single copy of the state, it will be possible given sufficiently many copies of the state.}
\EQ{
\overline{f(\rho_A(W),\rho_A)}&=\lim_{n\to\frac12}\tr^{(2n)} \big(\rho_{F_q}^\text{sc}(W)\otimes\rho_{F_q}^{\text{sc}}\big)^{\otimes n}\\[5pt]
&= \sqrt{\tr\rho_{F_q}^\text{sc}(W)\,\tr\rho_{F_q}^\text{sc}}=1\ ,
\label{tom2}
}
which is another expression of decoupling. Note that, if the trace norm difference of two states vanishes, then they must have unit quantum fidelity and $\rho_A(W)=\rho_A$.

On the other hand, when $F_q\in{\cal W}(A)$, the element $\tau_{q+1}=\eta$ and the trace norm difference \eqref{epi} is 
\EQ{
\overline{\NORM{\rho_A(W)-\rho_A}_1}&=\lim_{n\to\frac12}\tr^{(2n)}\eta\big(\rho_{F_q}^\text{sc}(W)-\rho_{F_q}^\text{sc}\big)^{\otimes2n}\\[5pt] &=\NORM{\rho_{F_q}^\text{sc}(W)-\rho_{F_q}^\text{sc}}_1\ .
\label{tum3}
}
For the fidelity, we have a similar relation to the semi-classical state
\EQ{
\overline{f(\rho_A(W),\rho_A)}&=\lim_{n\to\frac12}\tr^{(2n)} \eta\big(\rho_{F_q}^\text{sc}(W)\otimes\rho_{F_q}^{\text{sc}}\big)^{\otimes n}\\[5pt] &
= f(\rho_{F_q}^\text{sc}(W),\rho_{F_q}^\text{sc})\ .
\label{tom3}
}

Let us take stock of the results and, in particular, relate them to  the {\it state reconstruction formula\/} of \cite{Qi:2021sxb}. This states that if there are two microscopic states $\rho_A$ and $\sigma_A$ such that the semi-classical saddles that dominate $\tr(\rho_A^{2n})$ and $\tr(\sigma_A^{2n})$ are the same (and preserve the $\mathbb Z_n$ symmetry of the replicas) then\footnote{We have stated the formula in a slightly more general way to include the case when $A$ is any subset of the radiation plus the black hole rather than all the radiation as considered in  \cite{Qi:2021sxb}. The condition for $\mathbb Z_n$ symmetry is satisfied by our saddles which involve only the elements $e$ or $\eta$ of $S_n$.} 
\EQ{
\NORM{\rho_A-\sigma_A}_1=\NORM{\rho^\text{sc}_{{\cal W}(A)}-\sigma^\text{sc}_{{\cal W}(A)}}_1\ ,
\label{dis}
}
up to ${\mathscr O}(G)$ corrections, where ${\cal W}(A)$ is the entanglement wedge of $A$. The fact that the map $V$ preserves the trace norm difference is on the same footing as the preservation of the relative entropy \cite{Almheiri:2019qdq,Chen:2019iro,Jafferis:2015del}. 

To relate this to our analysis, we identify $\sigma_R=\rho_R(W)$. The saddles associated to $\tr(\rho_A^{2n})$ and $\tr(\sigma_A^{2n})$ are the ones that determine the R\'enyi entropies and are therefore associated to the set of elements $\tau_p$, $p=1,\ldots,N$. The fact that they both have the same saddle is ensured by the fact that $W$ only acts on a small subset of the infalling modes and so cannot alter the dominant saddle.

Let us consider our results for the case $A=R$. Before the Page time, ${\cal W}(R)=R$ and so $F_q\not\in{\cal W}(R)$ and the formula \eqref{dis} implies
\EQ{
\NORM{\rho_R(W)-\rho_R}_1=0\ ,
} 
which is the decoupling condition \eqref{tum2} with $A=R$. This means that $W$ can be reconstructed on $B$. On the other hand, after the Page time, the entanglement wedge ${\cal W}(R)=R\cup I(R)$, so $F_q\in{\cal W}(R)$, since the island $I(R)$ contains the outgoing and infalling modes $I(R)=\overline R\cup F$ since it lies very close behind the horizon. Hence, \eqref{dis} implies
\EQ{
\NORM{\rho_R(W)-\rho_R}_1=\NORM{\rho_{R\overline RF}^\text{sc}(W)-\rho_{R\overline RF}^\text{sc}}_1=\NORM{\rho_{F_q}^\text{sc}(W)-\rho_{F_q}^\text{sc}}_1\ ,
\label{dis4a}
}
which is \eqref{tum3} with $A=R$. We will see shortly that this is the case when $W$ can be reconstructed on $R$ because $B$ decouples.

Now consider the case $A=B$. After the Page time, ${\cal W}(B)=\varnothing$ and so \eqref{dis} predicts  decoupling as we found in \eqref{tum2}. This occurs at the same time as \eqref{dis4a} which makes perfect sense as $W$ can be reconstructed on $R$. On the other hand, before the Page time,  ${\cal W}(B)=\overline R\cup F$, and so \eqref{dis} gives
\EQ{
\NORM{\rho_B(W)-\rho_B}_1=\NORM{\rho_{\overline RF}^\text{sc}(W)-\rho_{\overline RF}^\text{sc}}_1=\NORM{\rho_{F_q}^\text{sc}(W)-\rho_{F_q}^\text{sc}}_1\ .
\label{dis3}
}
But this is precisely \eqref{tum3} for $A=B$. This is also when $R$ decouples and so $W$ can be reconstructed on $B$. So once again we find precise agreement between our averaged results and the formula \eqref{dis}.

\subsection{Bounding the trace norm}\label{s5.1}

The condition for decoupling is that the averaged trace norm difference between $\rho_A(W)$ and $\rho_A$ vanishes in the leading order saddle \eqref{tum2}. But this is derived with the limits in a particular order, first  the semi-classical limit picking out a particular saddle and then in the replica limit $n\to\frac12$. Can we trust this? In fact there is standard way to bound the averaged trace norm difference,
\EQ{
\overline{\NORM{\rho_A(W)-\rho_A}_1}\leq \sqrt{d_A\,\overline{\tr(\rho_A(W)-\rho_A)^2}}\ .
}
We can evaluate the right-hand side, at least in the case that the subsystem $A$ is finite dimensional. Note that this seems to exclude $A=R$, the radiation, in the refined model as it has infinite dimension. The average on the right-hand side is just the right-hand side of \eqref{bez} with $n\to1$, so
\EQ{
\overline{\tr(\rho_A(W)-\rho_A)^2}=\sum_{\tau_1,\ldots,\tau_N\subset\{e,\eta\}}e^{-S^{(2)}_{\{\tau_p\}}(A)}
\,\tr^{(2)}\tau_{q+1}\big(\rho_{F_q}^\text{sc}(W)-\rho_{F_q}^\text{sc}\big)^{\otimes2}\ .
\label{bez4}
}

If we consider $A=B$, so $d_A\sim e^{S_\text{BH}(M)}$, and after the Page time, the sum in \eqref{bez4} is dominated by the term with $\tau_p=\eta$ for which $S^{(2)}_{\{\eta\}}(B)=\alpha S_\text{rad}(R)$, where $\alpha=1$ for the basic model and $\alpha=\frac34$, for the refined model.\footnote{The latter follows from \eqref{wil} with $\tau_p=\eta$, $p=1,\ldots,N+1$ and $\sigma_p=e$ giving $d(\tau_{p+1},\tau_p)=0$ and $d(\sigma_p,\tau_p)=n-1$ giving $S^{(n)}_{\{\eta\}}(B)=(n+1)S_\text{rad}(R)/(2n)$. This is the R\'enyi entropy of the radiation (see appendix \ref{a2}) and then taking $n=2$ gives $\frac34S_\text{rad}(R)$.} Therefore we can bound the trace norm difference
\EQ{
\overline{\NORM{\rho_A(W)-\rho_A}_1}\lessapprox {\mathscr O}(e^{\frac12S_\text{BH}(M)-\frac\alpha2S_\text{rad}(R)})\ll 1\ ,
}
after the Page time when $S_\text{rad}(R)\gg S_\text{BH}(B)$.

\section{Reconstruction of the Hawking partners}\label{s6}

In this section, we consider reconstruction for the Hawking partners which semi-classically are behind the horizon and part of the black hole. This problem was considered in the static model of \cite{Akers:2022qdl}, which essentially represents the holographic map at a fixed time. Therefore, it differs from the dynamical models considered in this paper. In addition, we note that section 8.2 of \cite{Akers:2022qdl} addressed a different yet related problem concerning the ability to distinguish certain interior states.

Conceptually the discussion is very similar to the reconstruction of the infalling system in the last section but the technical details are rather different. The idea is to consider  a unitary operator on the Hawking partners $\overline R$ and ask if it is possible to reconstruct this on some $A\subset R\cup B$, i.e.
\EQ{
\ket{\Psi(W)}\overset{?}=W_{A}\ket{\Psi}\ .
\label{vet}
}
As in section \ref{s5} the condition for such a reconstruction is the decoupling condition for the complement
\EQ{
\rho_{\overline A}(W)=\rho_{\overline A}\ ,
}
which can be analysed by calculating the trace norm difference or quantum fidelity.

In order to proceed, it is useful to deploy the following trick. Exploiting the entanglement between $\overline R$ and $R$, we can write the action of $W$ on the semi-classical state as the action of an operator $\widetilde W$ on $R$:
\EQ{
W\ket{\psi}=\widetilde W\ket{\psi}\ ,
\label{cam}
}
where 
\EQ{
\widetilde W=(\rho^\text{sc}_R)^{1/2} W^T(\rho^\text{sc}_R)^{-1/2}\ .
}
We remark that $\widetilde W$ is not unitary so it is not a physically realizable local action on the radiation.
It then follows that the reduced state on $R$ is invariant under adjoint action by $\widetilde W$,
\EQ{
\rho^\text{sc}_R\longrightarrow\widetilde W\rho^\text{sc}_R\widetilde W^\dagger = (\rho^\text{sc}_R)^{1/2}\big(W^\dagger W\big)^*(\rho^\text{sc}_R)^{1/2}=\rho^\text{sc}_R\ ,
\label{mom}
} 
as it must be by locality: the action of $W$ on $\overline R$ cannot change the state of $R$.

We now compute the trace difference and quantum fidelity of the two states $\rho_A(W)$ and $\rho_A$ using the replica method following the same steps as in section \ref{s5}. For simplicity, we will take $W$ to act on just one of the subsets of partner modes $\overline R_q$. We can then use \eqref{cam} to write the action on the Hawking modes $R_q$ by switching $W\to\widetilde W$. As for the infalling system, the only effect of $W$ is on the $q^\text{th}$ time step. For the trace norm difference, this time step is modified as 
\EQ{
{\cal Z}_q&\longrightarrow \tr^{(2n)}\sigma_q^{[R_q]}\tau_{q+1}^{[B_q]}\big(\text{Ad}_{\widetilde W}-1\big)^{\otimes2n}(\tau_q^{[R_qB_q]})^{-1}\rho^{\text{(mm)}\ \otimes2n}_{R_qB_q}\ ,
}
where $\text{Ad}_{\widetilde W}$ is the adjoint action of $\widetilde W$ on $\rho^{\text{(mm)}}_{R_qB_q}$.
We now assume that the saddle that dominates the entropy, and therefore the trace norm difference, has $\tau_{q+1}=\tau_q$. This means that $\overline R_q$ is not just before a QES. One can view this as avoiding an edge effect created by having a discrete model. In that case, we can perform the trace over $B_q$ to give the semi-classical state $\rho^\text{sc}_{R_q}=\tr_{B_q}\rho_{R_qB_q}^\text{(mm)}$:
\EQ{
{\cal Z}_q\longrightarrow \tr^{(2n)}\sigma_q\big(\text{Ad}_{\widetilde W}-1\big)^{\otimes2n}\tau_q^{-1}\,\rho^{\text{sc}\ \otimes2n}_{R_q}\ .
\label{yet3}
}
where in the second line we used the fact that all relevant saddles have $\tau_{q+1}=\tau_q$ and $\rho^\text{sc}_{R_q}=\tr_{B_q}\rho_{R_qB_q}^\text{(mm)}$. Following the same steps as in section \ref{s5}, and in particular taking the semi-classical limit before the analytic continuation in $n$, gives 
\EQ{
\boxed{\overline{\NORM{\rho_A(W)-\rho_A}_1}=\lim_{n\to\frac12}\tr^{(2n)}\sigma_q\big(\text{Ad}_{\widetilde W}-1\big)^{\otimes2n}\tau_q^{-1}\,\rho^{\text{sc}\ \otimes2n}_{R_q}}
\label{rupa}
}
where $\tau_q$ is determined by the saddle that dominates the entropy.
Similarly, for the quantum fidelity
\EQ{
\boxed{\overline{f(\rho_A(W),\rho_A)}=\lim_{n\to\frac12}\tr^{(2n)}\sigma_q\big(
\text{Ad}_{\widetilde W}\otimes 1\big)^{\otimes n}\tau_q^{-1}\,
\rho^{\text{sc}\ \otimes 2n}_{R_q}}
}

When $\overline R_q\not\in{\cal W}(A)$ the dominant saddle has $\tau_q=e$ and then the trace norm difference is
\EQ{
\overline{\NORM{\rho_A(W)-\rho_A}_1}=\lim_{n\to\frac12}\tr^{(2n)}\sigma_q\big(\widetilde W\rho^\text{sc}_{R_q}\widetilde W^\dagger-\rho^\text{sc}_{R_q}\big)^{2n}=0\ ,
\label{rup}
}
using the invariance \eqref{mom}. We can repeat the analysis for the fidelity,
\EQ{
\overline{f(\rho_A(W),\rho_A)}&=\lim_{n\to\frac12}\tr^{(2n)}\sigma_q\Big(\widetilde W\rho^\text{sc}_{R_q}\widetilde W^\dagger\otimes\rho^{\text{sc}}_{R_q}\Big)^{\otimes n}\\[5pt]
&=\lim_{n\to\frac12}\tr^{(2n)}\sigma_q\,\rho^{\text{sc}\ \otimes2n}_{R_q}=1\ .
}
So decoupling occurs when the partners $\overline R_q$ do not lie in the entanglement wedge of $A$. Under these circumstances, $W$ can be reconstructed on the complement $\overline A$.

On the other hand, when $\overline R_q\in{\cal W}(A)$, the appropriate saddle has $\tau_q=\eta$ and \eqref{rupa} becomes
\EQ{
\overline{\NORM{\rho_A(W)-\rho_A}_1}=\lim_{n\to\frac12}\tr^{(2n)}\sigma_q\big(\text{Ad}_{\widetilde W}-1\big)^{\otimes2n}\eta^{-1}\rho^{\text{sc}\ \otimes2n}_{R_q}\ .
\label{rup2}
}
We can now consider this for particular choices for $A$. For the case $A=R$, so after the Page time, then $\sigma_q=\eta$, and the above becomes 
\EQ{
\overline{\NORM{\rho_R(W)-\rho_R}_1}&=2\sqrt{1-\big|\tr\big(\rho^\text{sc}_{R_q}W^T\big)\big|^2}\\[5pt]
&=\NORM{\rho_{R\overline R}^\text{sc}(W)-\rho_{R\overline R}^\text{sc}}_1\ .
\label{kee}
}
For the case $A=B$, so before the Page time, $\sigma_q=e$, we have
\EQ{
\overline{\NORM{\rho_B(W)-\rho_B}_1}&=\lim_{n\to\frac12}\tr\big(W^*\rho^\text{sc}_{R_q}W^T-\rho^\text{sc}_{R_q}\big)^{2n}\\[5pt]
&=\lim_{n\to\frac12}\tr\big(W\rho^\text{sc}_{\overline R_q}W^\dagger-\rho^\text{sc}_{\overline R_q}\big)^{2n}\\[5pt]
&=\NORM{\rho^\text{sc}_{\overline R}(W)-\rho^\text{sc}_{\overline R}}_1\ .
\label{kut}
}
Note that \eqref{kut} is not the same as \eqref{kee} because $R$ is entangled with $\overline R$.

We can also consider the quantum fidelity. For $A=R$ (after the Page time),
\EQ{
\overline{f(\rho_R(W),\rho_R)}&=\lim_{n\to\frac12}\tr^{(2n)}\eta
\big(\text{Ad}_{\widetilde W}\otimes 1\big)^{\otimes2n}\eta^{-1}
\rho^{\text{sc}\ \otimes 2n}_{R_q}\\[5pt]
&=\big|\tr(\rho^\text{sc}_{R_q}W^T)\big|=f(\rho^\text{sc}_{R\overline R}(W),\rho^\text{sc}_{R\overline R})
}
and for $A=B$ (before the Page time),
\EQ{
\overline{f(\rho_B(W),\rho_B)}&=\lim_{n\to\frac12}\tr^{(2n)}
\big(\text{Ad}_{\widetilde W}\otimes 1\big)^{\otimes2n}\eta^{-1}
\rho^{\text{sc}\ \otimes 2n}_{R_q}\\[5pt]
&=\lim_{n\to\frac12}\tr^{(2n)}\eta\big(W^*\rho^\text{sc}_{R_q}W^T\otimes \rho_{R_q}^{\text{sc}}\big)^n\\[5pt]
&=f(\rho^\text{sc}_{\overline R}(W),\rho^\text{sc}_{\overline R})\ .
}
These expressions are close cousins of the expressions for the trace norm difference in \eqref{kee} and \eqref{kut}.

Once again, let us compare our results to the formula \eqref{dis} of \cite{Qi:2021sxb}. Firstly, let us compare the microscopic states $\rho_R(W)$ and $\rho_R$. Before the Page time, $\overline R_q\not\in{\cal W}(R)$ and so \eqref{dis} implies $\NORM{\rho_R(W)-\rho_R}_1=0$. After the Page time, $\overline R_q\in{\cal W}(R)$  and so \eqref{dis} implies
\EQ{
\NORM{\rho_R(W)-\rho_R}_1=\NORM{\rho_{R\overline RF}^\text{sc}(W)-\rho_{R\overline RF}^\text{sc}}_1\ ,
\label{dis3}
}
which is precisely \eqref{kee} because $F$ is not entangled with $R\cup\overline R$. 

Now we turn to the states $\rho_B(W)$ and $\rho_B$. In this case, after the Page time, ${\cal W}(B)=\varnothing$ and so \eqref{dis} implies $\NORM{\rho_B(W)-\rho_B}_1=0$. On the other hand, before the Page time, ${\cal W}(B)=\overline R\cup F$, and so \eqref{dis} implies
\EQ{
\NORM{\rho_R(W)-\rho_R}_1=\NORM{\rho_{\overline RF}^\text{sc}(W)-\rho_{\overline RF}^\text{sc}}_1\ .
\label{dis4}
}
This is precisely \eqref{kut} because $F$ is not entangled with $\overline R$.

\section{Discussion}\label{s7}

We have defined a simple model which accounts for energy conservation that captures the information flow of an evaporating black hole and have seen that many of the features of the model proposed in \cite{Akers:2022qdl} can be extended to this refined model. Unitarity is built in and this manifests at the level of the entropy of the radiation in the form of a discrete version of the QES variational problem. The model then allowed us to investigate in detail entanglement wedge reconstruction for a system that falls into the black hole and also for local actions on the Hawking partners. Our main contributions here were to study these problems in a model which generalised the model of \cite{Akers:2022qdl} and also to consider when reconstruction is possible not just on the radiation or the black hole, but a subset thereof. The model reproduces the properties of the holographic map that have been proposed in \cite{Akers:2022qdl}; namely, the map acts trivially on the outgoing radiation and non-isometrically on the black hole. This latter fact manifests the fact that the Hilbert space of an old black hole is not large enough to host all the Hawking partners of the semi-classical state. Something must give, the map is non-isometric and as a result the Hawking partners have been teleported out into the radiation as subtle features of the microscopic state of $R$. In a sense, when a black hole is past the Page time according to an external observer, its inside has been squeezed out into the radiation leaving only a small region between the horizon and the QES that could be thought of as being part of the black hole.

Although the proposal of \cite{Akers:2022qdl} has clarified certain issues, much remains to be understood. Of principal interest is the fate of an infalling system. According to these models, an infalling system begins to be scrambled immediately. In fact, the infalling system will soon enter the entanglement wedge of a late-time observer who collects all the radiation, since the QES is very close up behind the horizon meaning that the information of the infalling observer is in the radiation available to the late-time observer. Is this compatible with the idea that the infalling system experiences a smooth internal geometry after horizon crossing? We have argued at the microscopic level, the state of the radiation is not the inertial vacuum in the neighbourhood of the horizon but perhaps the infalling system sees effectively a smooth geometry and being thermalized takes some time. This would support an idea previously presented in \cite{Susskind:2015toa} and analysed in the basic model in \cite{Akers:2022qdl}. The situation seems quite analogous to the same questions for the fuzzball paradigm in string theory \cite{Mathur:2012np,Mathur:2012jk}. In that context, it is argued that a macroscopic (i.e.~high energy) infalling system would take time to be thermalized as it falls into the fuzzball. In a proposal known as {\it fuzzball complementarity\/}, the high energy infallling system would not resolve the subtle structure of the microscopic state and effectively average it to see a smooth geometry. It seems plausible that the same mechanism is at work here, if an observer cannot resolve the fine details of $\rho_R$ maybe it effectively experiences the average $\overline{\rho_R}=\rho^\text{sc}_R$, precisely the semi-classical state and a smooth horizon, at least for a while.

\vspace{0.5cm}
\begin{center}{\it Acknowledgments}\end{center}
\vspace{0.2cm}
TJH, AL and SPK acknowledge support from STFC grant ST/T000813/1. NT and ZG acknowledge the support of an STFC Studentship.
AL has also received funding from the European Research Council (ERC) under the European Union’s Horizon 2020 research and innovation programme (grant agreement No 804305).

\vspace{0.5cm}

\begin{center}
********************
\end{center}

\vspace{0.5cm}

{\small For the purpose of open access, the authors have applied a Creative Commons Attribution (CC BY) licence to any Author Accepted Manuscript version arising.}

\appendix
\appendixpage

\section{Thermodynamics of free fields}\label{a2}

Consider a set of free fields in $1+1$ dimensions. We will consider just the right-moving modes. The canonical partition function of a single mode of energy $\omega$ is equal to
\EQ{
{\cal Z}=\sum_{p=0}^\infty e^{-p\omega/T}=\frac1{1-e^{-\omega/T}}\ ,\qquad \sum_{p=0}^1 e^{-p\omega/T}=1+e^{-\omega T}\ ,
}
for a scalar and spinor field, respectively. Summing over modes in a volume $\mathscr V$ and assuming there are $N=c,2c$ fields for bosons/fermions, gives the free energy
\EQ{
f=\pm N\mathscr VT\int_0^\infty\frac{d\omega}{2\pi}\,\log(1\mp e^{-\omega/T})=-\frac{\pi c\mathscr VT^2}{12}\ .
\label{dud}
}
The average energy 
\EQ{
{\cal E}=N\mathscr V\int_0^\infty\frac{d\omega}{2\pi}\,\frac{\omega}{e^{\omega/T}\mp1}=\frac{\pi c\mathscr VT^2}{12}
\label{see}
}
and the entropy
\EQ{
S_\text{rad}=N\mathscr V\int_0^\infty\frac{d\omega}{2\pi}\,\Big\{\frac{\omega}{T(e^{\omega/T}\mp1)}\mp\log(1\mp e^{-\omega/T})\Big\}=\frac{\pi c\mathscr VT}6\ .
}

We can also evaluate the R\'enyi entropes,
\EQ{
(1-n)S_\text{rad}^{(n)}&=N\mathscr V\int_0^\infty\frac{d\omega}{2\pi}\,
\log\sum_{p=0}^{\infty,1}\Big(\frac{e^{-p\omega/T}}{\cal Z}\Big)^n\\[5pt] &=N\mathscr V\int_0^\infty\frac{d\omega}{2\pi}\,\big(\log{\cal Z}(T/n)-n\log{\cal Z}(T)\big)\ .
}
Hence,
\EQ{
S_\text{rad}^{(n)}=\frac{n f(T)-nf(T/n)}{(1-n)T}=\frac{1+n}n\,\mu T=\frac{1+n}{2n}\,S_\text{rad}\ .
}

We will need to understand whether the relativistic gas can be described thermodynamically. We can solve for the entropy in terms of the entropy, $S_\text{rad}=2\sqrt{\mu{\cal E}}$, where $\mu=\pi c\mathscr V/12$. In the thermodynamic it should be possible to approximate the canonical partition function as a integral over a continuum set of states with energy ${\cal E}$ and density of states $e^{S_\text{rad}({\cal E})}$, that is
\EQ{
{\cal Z}=e^{-f/T}=\int d{\cal E}\,e^{S_\text{rad}({\cal E})-{\cal E}/T}\ .
}
The thermodynamic limit can be understood as when the saddle point approximation of this integral is valid. The saddle point equation corresponds to the Legendre transformation between the internal energy and free energy:
\EQ{
f=\underset{\cal E}\ext\big({\cal E}-TS_\text{rad}({\cal E})\big)\ ,
}
and has solution
\EQ{
{\cal E}=\mu T^2\ ,
}
for which the free energy 
\EQ{
f=-\mu T^2\ .
}
One can verify that these expressions are entirely consistent with \eqref{dud} and \eqref{see}. The saddle point approximation is valid in the limit that the spread in the energy around the saddle point $\Delta{\cal E}\ll {\cal E}$ which is the condition
\EQ{
\frac{\Delta{\cal E}}{\cal E}\thicksim \frac1{\sqrt{S_\text{rad}}}\ll1\ .
}
So when $S_\text{rad}\gg1$, the gas can be described thermoydnamically.

\section{Dominant saddles}\label{a1}

In the model, we encounter sums over elements of the symmetric group of the form \eqref{ruc}. This motivates analysing a sum of the form
\begin{align} \label{sumzn}
{\cal Z}(n) = \sum_{\sigma \in S_n} d_1^{-d(\sigma,\tau_1)} d_2^{-d(\sigma,\tau_2)} d_3^{-d(\sigma,\tau_3)} \ ,
\end{align}
where $d_i \geq 1$, $\tau_i \in S_n$ and $d(\sigma,\pi)$ is the Cayley distance between elements of $S_n$. This is equal to
\EQ{
d(\sigma,\pi)=n-k(\sigma\pi^{-1})\ ,
}
where $k(\sigma)$ is the number of cycles the make up $\sigma$, e.g.~$k(e)=n$ and $k(\eta)=1$.

\begin{figure}[h]
\begin{center}
\begin{tikzpicture} [scale=1]
\draw[<->,thick,black!60] (0,4.5) -- (0,0) -- (4.5,0);
\draw[blue,thick] (4,2) -- (2,0) -- (0,2) -- (2,4);
\node at (5.2,0) {$x_1/x_3$};
\node at (0,5) {$x_2/x_3$};
\node at (2,-0.4) {$1$};
\node at (-0.7,2) {$1$};
\node[black] at (2,2) {$\Gamma(\tau_1,\tau_2,\tau_3)$};
\node[black] at (3.5,0.5) {$\tau_1$};
\node[black] at (0.5,3.5) {$\tau_2$};
\node[black] at (0.5,0.5) {$\tau_3$};
\end{tikzpicture}
\caption{\small Phase diagram for the sum \eqref{sumzn} when $\Gamma(\tau_1,\tau_2,\tau_3)$ is not empty. Along the blue lines there are more permutations which can contribute e.g. along $x_1/x_3 + x_2/x_3 = 1$ the sum is dominated by the set of permuations which lie in $\Gamma(\tau_1,\tau_3) \cap \Gamma(\tau_2,\tau_3)$.} \label{fig:phasediagram}
\end{center}
\end{figure}
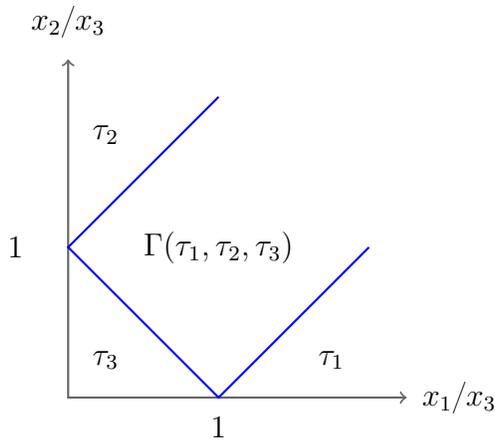

We are interested in minimising the following `free energy'
\begin{align}
f(\sigma) = x_1 d(\sigma,\tau_1) + x_2 d(\sigma,\tau_2) + x_3 d(\sigma,\tau_3) \ ,
\end{align}
where $x_i = \log d_i$. We first consider the permutations which minimise the free energy at the following special regions in the {\it phase diagram\/} (see figure \ref{fig:phasediagram}), which we may parameterise by $x_1/x_3$ and $x_2/x_3$:
\begin{itemize}
\item for $x_1/x_3 \to 0$ and $x_2/x_3 \to 0$: $f(\sigma) \to x_3 d(\sigma, \tau_3)$ is minimised for $\sigma = \tau_3$.
\item for $x_1/x_3 + x_2/x_3 = 1$: $f(\sigma) = x_1 \left(d(\tau_1,\sigma)+d(\sigma,\tau_3)\right) + x_2 \left(d(\tau_2, \sigma)+d(\sigma,\tau_3)\right)$ is minimised for $\sigma \in \Gamma(\tau_1, \tau_3) \cap \Gamma(\tau_2,\tau_3)$. Here, $\Gamma(\tau_i, \tau_j)$ denotes the set of permutations $\sigma$ which saturate the triangle inequality $d(\tau_i,\sigma) + d(\sigma,\tau_j) \geq d(\tau_i,\tau_j)$.
\end{itemize}
There are two $+$ two more regions in the phase diagram where the permutations which minimise the free energy can be determined by cyclically permuting the labels in the above. Most of the rest of the phase diagram can then be filled in using convexity of the free energy. That is, since $f$ is a linear function of the $x_i$, if $\sigma$ minimises $f$ at two points in the phase diagram, then $\sigma$ also minimises $f$ along the segment joining these two points. This argument can only be used to fill in the whole phase diagram if the set of permutations $\Gamma(\tau_1,\tau_2,\tau_3) \coloneq \Gamma(\tau_1,\tau_2) \cap \Gamma(\tau_2,\tau_3) \cap \Gamma(\tau_3,\tau_1)$ which simultaneously saturate the three triangle inequalities
\begin{align}
d(\tau_i,\sigma)+d(\sigma,\tau_j) \geq d(\tau_i,\tau_j) \quad \text{for $i \neq j$} \ ,
\end{align}
is not empty. The argument we have used to find the minima of $f$ by considering special regions in the phase diagram and then using convexity to fill in the rest is due to \cite{Akers:2021pvd}. 

From the above, we find that:
\begin{itemize}
\item for $x_1/x_3 + x_2/x_3 < 1$:
\begin{align} \label{lemma 1}
{\cal Z}(n) \approx d_1^{-d(\tau_1,\tau_3)} d_2^{-d(\tau_2,\tau_3)} \ .
\end{align}
The behaviour of the sum in two other regions may be obtained by cyclically permuting the labels in the above.
\item assuming $\Gamma(\tau_1,\tau_2,\tau_3)$ is not empty, for $x_1/x_3 + x_2/x_3 > 1$, $x_2/x_1 + x_3/x_1 > 1$ and $x_3/x_2 + x_1/x_2 > 1$:
\begin{align} \label{lemma 2}
{\cal Z}(n) \approx |\Gamma(\tau_1,\tau_2,\tau_3)| \Big(\frac{d_1d_2}{d_3}\Big)^{-d(\tau_1,\tau_2)/2} \Big(\frac{d_2d_3}{d_1}\Big)^{-d(\tau_2,\tau_3)/2} \Big(\frac{d_3d_1}{d_2}\Big)^{-d(\tau_3,\tau_1)/2} \ .
\end{align}
\end{itemize}

\subsection{Proof}

We now prove that when $\tau_N \in \{e, \eta\}$ the nested sum \eqref{ruc} in the simple model:
\begin{align}
{\cal Z}_N(\tau_N) = \sum_{\tau_0, \dots, \tau_{N-1} \in S_n} \prod_{p=1}^N d_{B_p}^{-d(\tau_{p-1},\tau_p)} d_{R_p}^{-d(\tau_{p-1},\sigma_p)} \ ,
\end{align}
with $d_{B_p}, d_{R_p} \geq 1$ and $\sigma_p \in \left\{e,\eta\right\}$, is dominated by the terms with $\tau_{p-1} \in \{e,\eta\}$ for each $1 \leq p\leq N$, provided we ignore the crossover regimes. It is useful to notice that ${\cal Z}_N(\tau_N)$ satisfies the recursion relation
\begin{align}
{\cal Z}_N(\tau_N) = \sum_{\tau_{N-1} \in S_n} d_{B_N}^{-d(\tau_{N-1}, \tau_N)} d_{R_N}^{-d(\tau_{N-1}, \sigma_N)}{\cal Z}_{N-1}(\tau_{N-1})\ , \qquad{\cal Z}_0(\tau_0) = 1 \ .
\end{align}
First consider
\begin{align}
{\cal Z}_1(\tau_1) = \sum_{\tau_0 \in S_n} d_{B_1}^{-d(\tau_0,\tau_1)} d_{R_1}^{-d(\tau_0,\sigma_1)} \ .
\end{align}
This sum is of the form \eqref{sumzn} so is dominated by the terms with $\tau_0 \in \{\sigma_1,\tau_1\} \subset \{e,\eta,\tau_1\}$. Using this fact we see that
\begin{equation}
\begin{aligned}
{\cal Z}_2(\tau_2) &= \sum_{\tau_1 \in S_n} d_{B_2}^{-d(\tau_1,\tau_2)} d_{R_2}^{-d(\tau_1,\sigma_2)} {\cal Z}_1(\tau_1)\\
&\approx\sum_{\tau_1 \in S_n} d_{B_2}^{-d(\tau_1,\tau_2)} d_{R_2}^{-d(\tau_1,\sigma_2)} \min(d_{B_1},d_{R_1})^{-d(\tau_1,\sigma_1)} \ ,
\end{aligned}
\end{equation}
is also of the form \eqref{sumzn} so is dominated by the terms with $\tau_1 \in \{\sigma_1,\sigma_2,\tau_2\} \cup \Gamma(\sigma_1,\sigma_2,\tau_2) \subset \{e,\eta,\tau_2\} \cup \Gamma(e,\eta,\tau_2)$. We have assumed that $\Gamma(e,\eta,\tau_2)$ is not empty; a fact we will verify ex-post facto. Using this, \eqref{lemma 1} and \eqref{lemma 2} it is simple to show that ${\cal Z}_3(\tau_3)$ is also of the form \eqref{sumzn} so is dominated by the terms with $\tau_2 \in \{e,\eta,\tau_3\} \cup \Gamma(e,\eta,\tau_3)$.\footnote{There is a slight subtlety here as the sum ${\cal Z}_3(\tau_3)$ can differ from \eqref{sumzn} by a factor of $|\Gamma(\sigma_1,\sigma_2,\tau_2)|$. Whilst this factor depends on $n$, it is independent of $d_{B_p}$ and $d_{R_p}$ so it is reasonable to expect that we can ignore its effect if we are interested in the limit where $d_{B_p}$ and $d_{R_p}$ are large and eventually also the limit $n \to 1$.} Again, we have assumed that $\Gamma(e,\eta,\tau_3)$ is not empty; a fact we will verify ex-post facto. It is not too difficult to see that this pattern continues and proceeding with the argument we find that, provided $\Gamma(e,\eta,\tau_p)$ is not empty,
\begin{align}
\tau_{p-1} \in \{ e, \eta, \tau_p \} \cup \Gamma(e,\eta,\tau_p)
\end{align}
for each $1 \leq p \leq N$. However, since $\tau_N \in \{e,\eta\}$, this implies that
\begin{align}
\tau_{p-1} \in \{e,\eta\}
\end{align}
for each $1 \leq p \leq N$. In particular, each $\Gamma(e,\eta,\tau_p)$ is not empty, which is consistent with our assumption.

\end{document}